\documentclass[useAMS,usenatbib,usegraphicx]{mn2e}
\usepackage{times}

\begin{document}

\newcommand{\be}{\begin{equation}}
\newcommand{\ee}{\end{equation}}
\newcommand{\dl}{D_{\rm l}}
\newcommand{\ds}{D_{\rm s}}
\newcommand{\te}{t_{\rm E}}
\newcommand{\dand}{D_{\rm M31}}
\newcommand{\sm}{{\rm M}_{\sun}}
\newcommand{\den}{{\rm M}_{\sun}\,\mbox{pc}^{-3}}

\title[Impact of spheroid stars for microlensing in Andromeda]{The
impact of spheroid stars for Macho microlensing surveys of the
Andromeda Galaxy}
\author[Kerins]{Eamonn Kerins$^1$\\
$^1$Astrophysics Research Institute, Liverpool John Moores
 University, 12 Quays House, Birkenhead, Merseyside CH41 1LD.}
\maketitle

\begin{abstract}
The Andromeda Galaxy (M31) is an important test case for a number of
microlensing surveys looking for massive compact halo objects
(Machos). A long-standing theoretical prediction is that the high
inclination of the M31 disk should induce an asymmetry in the spatial
distribution of M31 Macho events, whilst the distribution of variable
stars and microlensing events in the M31 disk should be symmetric. We
examine the role of stars in the M31 visible spheroid as both lenses
and sources to microlensing events. We compute microlensing event
number density maps and estimate pixel-lensing rates and event
durations for three-component models of M31 which are consistent with
the observed rotation curve, surface brightness profile and dynamical
mass estimates. Three extreme models are considered: a massive
spheroid model; a massive disk model; and a massive Macho halo
model. An important consequence of the spheroid is that, even if
Machos are absent in M31, an asymmetric spatial signature is still
generally expected from stellar lensing alone. The lensing of disk
sources by spheroid stars produces an asymmetry of the same sign as
that of Machos, whilst lensing by disk stars against spheroid sources
produces an asymmetry of opposite sign. The relative mass-to-light
ratio of the spheroid and disk populations controls which of these
signatures dominates the overall stellar spatial distribution. We find
that the inclusion of the spheroid weakens the M31 Macho spatial
asymmetry by about $20-30\%$ over a disk-only asymmetry for the models
considered. We also find for our models that Machos dominate
over most of the far disk provided they contribute at least $\sim
25\%$ of the halo dark matter density. This is a conservative limit
since many stellar events are too short to be detected by present
surveys. The presence of the spheroid also has beneficial consequences
for M31 lensing surveys. The stellar spatial asymmetry is likely to be
important in distinguishing between a spheroidal Macho halo or a
highly flattened halo or dark matter dominated disk, since spatial asymmetries
of opposing signs are expected in these cases.
\end{abstract}


\section{Introduction}

In recent years the Andromeda Galaxy (M31) has become a key target for
microlensing experiments which are trying to detect massive compact
halo objects, or Machos (\citealt{aur01,cro01,riff01}). There are two
reasons for this. The first is the development of new techniques which
allow microlensing to be detected reliably against unresolved stellar
fields, enabling microlensing searches to be directed toward more
distant galaxies. The second reason is that the evidence gathered by
the first generation of microlensing experiments towards the
Magellanic Clouds remains inconclusive regarding the existence of
Machos (\citealt{alc00,las00,afo03}). This is largely because the
contribution of the Magellanic Clouds themselves to the observed
microlensing rate is highly uncertain.

The favourably high inclination of the M31 disk $(i = 77^\circ)$
presents a very promising diagnostic for the current experiments to
establish the presence or absence of Machos.
\citet{cro92} noted that if M31 Machos have a spheroidal distribution
the inclined disk should induce a spatial asymmetry in the
distribution of M31 Macho microlensing events, with more events being
seen towards the far disk which lies behind a larger fraction of the
halo column density. By contrast, foreground Milky Way Machos, stellar
microlensing events in the M31 disk, and variable stars mistaken
for microlensing, are expected to be symmetrically distributed. The near-far
asymmetry prediction therefore provides, in principle, a clean and
simple way to distinguish M31 Machos from other populations. Several
theoretical studies have exploited this to establish how Macho or galactic
model parameters could be measured by the M31 microlensing surveys
(\citealt{gyuk00,ker01,bal03,ker03}).

The visible spheroid of M31 has the potential to complicate this
simple picture. Despite superficial similarities between our own
Galaxy and M31, their visible spheroid populations appear to be
different in two key respects. Firstly, several studies indicate that,
as well as comprising metal-poor stars, the M31 spheroid contains a
dominant metal-rich population (\citealt{hol96,dur01,bel03}), in
contrast to the exclusively metal-poor Galactic spheroid. Secondly,
the M31 spheroid has a much higher luminosity density than that of our
own Galaxy in relative terms \citep{rei98}. These two features may
have a common link. \citet{fer02} have reported the existence of
significant stellar substructure in the outer regions of
M31. Substructure is evident both as coherent stellar density
enhancements and as metallicity variations. The alignment of a stellar
stream to the location of the satellite galaxy M32 indicates that this
galaxy may be the origin of some of this substructure. \citet{fer02}
suggest that M32 may have once been a much larger galaxy and that its
material has been stripped to pollute and beef-up the M31 spheroid.

Whatever is the origin of the M31 spheroid, its impact on the expected
microlensing signature must be assessed. \cite{ker03} have already
noted that the low surface brightness stellar streams are unlikely to have a
significant impact for microlensing studies. The same cannot be
assumed for the M31 spheroid. Its shape and density means that it has
the potential to induce an asymmetry similar to that for M31 Machos,
making the task of identifying a Macho population more difficult. Its
stars may also act as sources to microlensing events arising from
other lens populations, modifying their spatial distribution.

In this study we investigate the likely effect of the
spheroid population on the spatial microlensing signature towards
M31. Section~\ref{asymm} provides analytical arguments for asymmetry
from stellar lenses. In section~\ref{models} we present a set of
extreme but simple three-component galactic models for M31 which are
likely to bracket realistic models. We compute microlensing event
number density maps and discuss microlensing rates for these models in
Section \ref{tau} and discuss the effect of the spheroid on Macho searches
in Section~\ref{discuss}.

\section{Asymmetry from stellar lenses} \label{asymm}

The predicted near-far asymmetry due to M31 Machos is expected to
occur because of the larger halo column density lying in front of the
far disk compared to that in front of the near disk \citep{cro92}. We
may therefore anticipate a similar signature due to M31 spheroid stars
acting as lenses against disk sources. However, in this case there is
also a reverse asymmetry due to lenses in the disk microlensing
background spheroid stars. Which of these two asymmetry signatures
might we expect to dominate the spatial signature of stellar lensing
events?  We can answer this question by considering the number density
of stellar lensing events in the two components. The event number
density along a given line of sight scales with lens (source) distance
$\dl$ ($\ds$) as
   \be
	d^2N \propto \sqrt{\frac{\dl(\ds - \dl )}{\ds} }
        \overline{v} \rho_{\rm l}(\dl ) \, d \dl  \rho_{\rm s}( \ds )
        \left(
	\frac{M}{L} \right)_{\rm s}^{-1} \, d\ds, \label{nev}
   \ee
where $\rho_{\rm l}$ ($\rho_{\rm s}$) is the lens (source) mass
density, $(M/L)_{\rm s}$ is the mass-to-light ratio of the source
population and $\overline{v}$ is the mean pairwise velocity of lenses and sources. Equation~(\ref{nev}) is proportional to
the rate of microlensing for sources at $\ds$, multiplied by the source
number density.

For a fixed source luminosity density the number of sources depends
upon the typical source luminosity $L_{\rm s}$. However, the expected
number of pixel-lensing events turns out to be independent of source
luminosity to first order.  The faintest detectable pixel-lensing
events require a minimum source magnification $A_{\rm min} \propto
L_{\rm s}^{-1}$ in order to be seen against the local surface
brightness flux. In pixel lensing the large bulk of observed events involve highly
magnified sources, for which $A_{\rm min} \simeq u_{\rm max}^{-1}$ and $u_{\rm
max}$ is the largest impact parameter between lens and source which
still provides a detectable event.  Since the pixel-lensing rate per
source $\Gamma_{\rm p} \propto u_{\rm max}$ it follows that
$\Gamma_{\rm p} \propto L_{\rm s}$. So, even though the number of
potential sources $N_{\rm s} \propto L_{\rm s}^{-1}$ for a fixed
surface brightness, the total number of pixel-lensing events $N =
N_{\rm s} \Gamma_{\rm p}$ does not depend on $L_{\rm s}$ to first
order.

For lines of sight passing through the major axis of M31 the disk
stars are, to a good approximation, at a fixed distance $\dand$ from
the observer. So, in the case of spheroid lenses and disk sources we
have $\rho_{\rm s} dD_{\rm s} = \Sigma_{\rm disk} \delta(\ds -
\dand)$, where $\Sigma_{\rm disk}$ is the disk column density and
$\delta(x)$ is the Dirac delta function, $\rho_{\rm l} = \rho_{\rm
sph}$ is the lens density distribution and $(M/L)_{\rm s} = (M/L)_{\rm
disk}$ is the source star mass-to-light ratio. The event number
density for spheroid lenses and disk sources for lines of sight
passing through the M31 major axis is therefore
   \be
	dN_{\rm sph} \propto \Sigma_{\rm disk} \left( 
        \frac{M}{L} \right)^{-1}_{\rm disk} \sqrt{ \frac{\dl (\dand - \dl)}
        {\dand} } \rho_{\rm sph}
        (\dl ) \, d\dl. \label{nevsph}
   \ee
In the case of disk lenses and spheroid sources we have $\rho_{\rm s}
= \rho_{\rm sph}$, $\rho_{\rm l} dD_{\rm l} = \Sigma_{\rm disk} \delta(\dand -
\dl)$ and $(M/L)_{\rm s} = (M/L)_{\rm sph}$, implying an event number
density
   \be
	dN_{\rm disk} \propto \Sigma_{\rm disk} \left(
	\frac{M}{L} \right)^{-1}_{\rm sph} \sqrt{ \frac{\dand(\ds -
	\dand)}{\ds} } \rho_{\rm sph} (\ds ) \, d\ds. \label{nevdisk}
   \ee
In both equation~(\ref{nevsph}) and (\ref{nevdisk}) above we have
dropped the factor $\overline{v}$ since the symmetry between the two
lens populations we are considering permits us to treat it as a
constant. Given the spatial symmetry of the disk and spheroid
populations for lines of sight passing through the M31 major axis we
can compare directly the rate contribution $dN_{\rm sph}$ for
spheroid lenses at $\dl = \dand - z$ to the contribution $dN_{\rm
disk}$ for sources at $\ds =
\dand + z$:
   \begin{eqnarray}
	\frac{dN_{\rm sph}}{dN_{\rm disk}} & = &
        \frac{1}{\dand}
        \frac{(M/L)_{\rm sph}}{(M/L)_{\rm disk}} \sqrt{(\dand - z)(\dand+z)}
	\nonumber \\
	& \simeq & \frac{(M/L)_{\rm sph}}{(M/L)_{\rm disk}} \quad \quad
        (z \ll \dand). \label{nevrat}
   \end{eqnarray}
This simple formula tells us that the direction of asymmetry for
stellar lenses is controlled by the mass-to-light ratio of
the disk and spheroid populations. If $M/L$ for the spheroid exceeds
that for the disk then one expects spheroid lenses to dominate number
counts on the major axis, and so the overall stellar asymmetry should
be in the same direction as that of M31 Machos, i.e. more events
towards the far disk. If the disk has a larger $M/L$ then disk lenses
should dominate and we can expect the stellar asymmetry to run counter to
that of M31 Machos.

\section{Extreme models of Andromeda}\label{models}

\begin{table*}
\begin{minipage}{160mm}
\caption{Parameters for the three extreme M31 models. From the left
the columns give, respectively, the model name, component population,
mass of the component, its central density $\rho_0$, and the cut-off
radius $R$. Additional columns give, where appropriate, the core
radius $a$, the $B$-band mass-to-light ratio $M/L_B$ in solar units,
the disk scale length $h$ and height $H$, and the flattening parameter
$q$. Parameters for a Reference model, representing a less extreme ``mainstream'' model are also given.}
\label{tab-param}
\begin{tabular}{@{}llcccccccc}
\hline
Model & Component & Mass $(\times 10^{10}\, \sm)$ & $\rho_0 (\den)$
& $R (\mbox{kpc})$ & $a (\mbox{kpc})$ & $M/L_B$ & $h (\mbox{kpc})$ &
$H (\mbox{kpc})$ & $q$ \\
\hline
Massive halo & halo & 191 & 0.25 & 155 & 2 & -- & -- & -- & 1 \\
 & spheroid & 4.4 & 4.5 &  40 & 1 & 9 & -- & -- & 0.6 \\
 & disk & 3.2 & 0.24 & 40 & -- & 4.5 & 6 & 0.3 & -- \\
Massive spheroid & halo & 89 & 0.01 & 85 & 10 & -- & -- & -- & 1 \\
 & spheroid & 8 & 2.5 & 40 & 1.5 & 14 & -- & -- & 0.6 \\
 & disk & 11 & 0.6 & 40 & -- & 18 & 7 & 0.3 & -- \\
Massive disk & halo & 79 & 0.01 & 110 & 8 & -- & -- & -- & 1 \\
 & spheroid & 4.4 & 4.5 & 40 & 1 & 9 & -- & -- & 0.6 \\
 & disk & 19 & 1.4 & 40 & -- & 26 & 6 & 0.3 & -- \\
\hline
Reference & halo & 123 & 0.065 & 100 & 4 & -- & -- & -- & 1 \\
 & spheroid & 4.4 & 4.5 & 40 & 1 & 9 & -- & -- & 0.6 \\
 & disk & 5.3 & 0.35 & 40 & -- & 8 & 6.4 & 0.3 & -- \\
\hline
\end{tabular}
\end{minipage}
\end{table*}

In assessing the likely contribution of the visible spheroid
population we choose to set up three extreme models which are likely
to bracket realistic models. We are not trying to present the models
as likely descriptions of the M31 structure, rather we wish to ensure
that the real signature lies somewhere within
the range of results we find for our extreme models.  Each of our
models comprises three components: a visible spheroid, a disk
population, and a dark halo. We do not include any nuclear bulge component
mainly because the bulge region is unlikely to be useful to
experiments wishing to measure any near-far spatial asymmetry arising
from Machos, but also because not including it allows us to test the
widest possible mass range for the other components.

Constraints on possible models come from the M31 surface brightness
profile, the rotation curve and from dynamical mass estimates, so we
ensure that our models satisfy at least these constraints. We also
ensure that the $B$-band $M/L$ values of the visible populations
in our models is not lower than considered reasonable for those
populations, though we do not impose any upper bound on $M/L_B$.

For our halo we assume a simple cored near-isothermal sphere, with a
density falling off with radius $r$ as
   \be
	\rho_{\rm halo} =\frac{\rho_0 }{1 + (r/a)^2}, \label{halo}
   \ee
where $\rho_0$ is the central density and $a$ is the core radius.
The disk is assumed to have a sech-squared distribution:
   \be
	\rho_{\rm disk} = \rho_0 \exp \left( -\frac{\sigma}{h} \right)
        {\rm sech}^2 \left( \frac{z}{H} \right), \label{disk}
   \ee
where $(\sigma,z)$ are cylindrical coordinates, $h$ the disk scale
length and $H$ the scale height.
Finally, we model the spheroid by a flattened power law:
   \be
	\rho_{\rm sph} = \frac{\rho_0}{1 + [(\sigma/a)^2 +
        q^{-2}(z/a)^2]^{s/2}}, \label{spheroid}
   \ee
with $q$ the ratio of minor to major axes and $s$ the power law
index. \citet{rei98} find that $s \simeq 3.8$ and $q \simeq 0.6$
provide a good fit to their number counts of RGB stars at a projected
distance $R = 19$~kpc from the M31 centre, together with number counts
from \citet{hol96} in fields at $R = 7.6$ and 10.8~kpc. However, a
single power law does not provide a good global description of the M31
spheroid, which is better fit by a de~Vaucouleurs profile \citep{pri94}.  However,
as we shall see, this simple parameterisation reproduces the surface
brightness distribution remarkably well in the inner galaxy. The main
effect for microlensing calculations in using a single power law
rather than a de~Vaucouleurs profile is to increase the spatial
dispersion of spheroid stars along the line of sight and so increase
their contribution to the microlensing rate. For such a steep profile
this is a small effect and, in any case, provides for a worst-case
assessment of the impact of spheroid stars for Macho microlensing
surveys.

The parameters of the galactic components are listed in
Table~\ref{tab-param} for the three extreme models. The massive halo
model assumes an extreme total M31 mass of around $2 \times
10^{12}\,\sm$, consistent with traditional estimates, whilst the other
two models assume a total mass closer to $1\times 10^{12}\,\sm$, in
line with recent determinations (\citealt{evans00,guh00}). The $M/L_B$
values for the disk and spheroid in the massive halo model are close
to minimal and do not allow much room for non-luminous matter. By
contrast in the massive spheroid model the disk and spheroid have
larger $M/L_B$ values. This model possesses a halo with a large core
radius and a disk with a slightly larger scale length in order to
increase the spheroid contribution. In the massive disk and spheroid
models the high disk $M/L_B$ assumes a dark matter dominated disk,
something which is known not to be the case for the Milky Way disk
\citep{cre98}. The disk and spheroid $M/L_B$ values are comparable for
the massive spheroid model, whilst the spheroid $M/L_B$ is higher than
that of the disk for the massive halo model and lower for the massive
disk model. We have also included a reference model in Table~\ref{tab-param}. This model employs a less extreme set of parameters and therefore can be considered a more likely candidate model than the extreme models.

\begin{figure*}
\includegraphics[scale=0.3,angle=270]{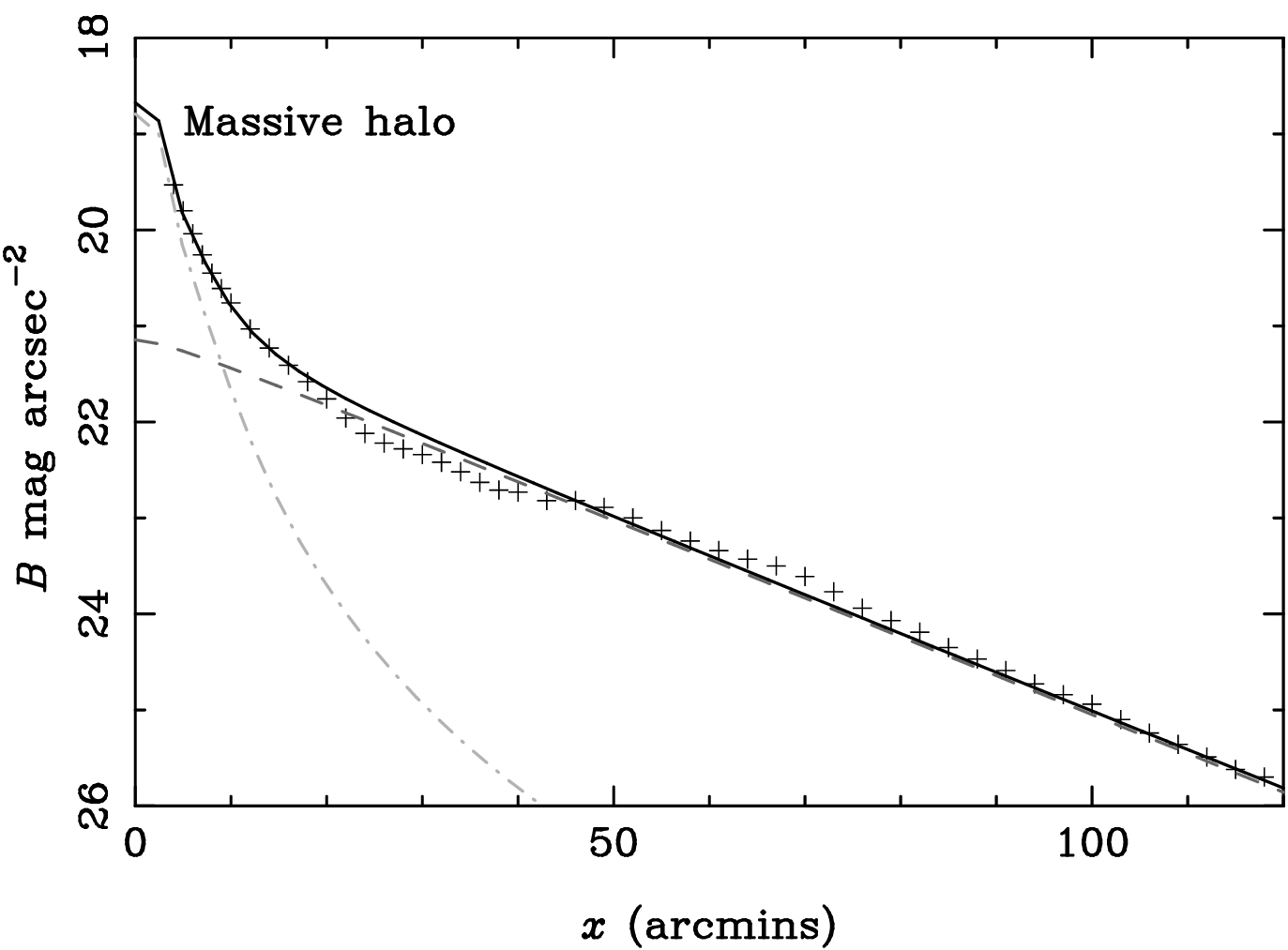}
\includegraphics[scale=0.3,angle=270]{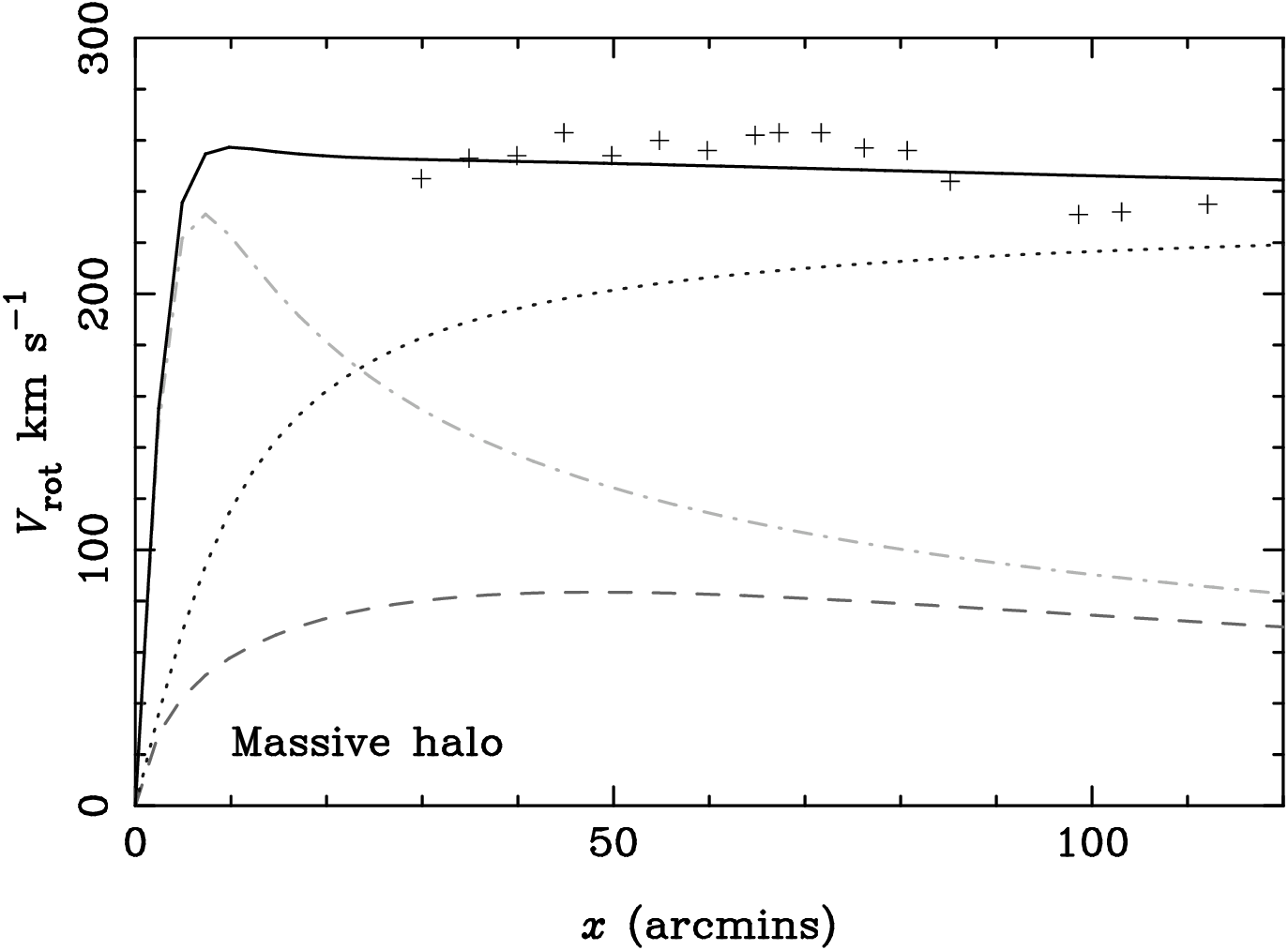}
\includegraphics[scale=0.3,angle=270]{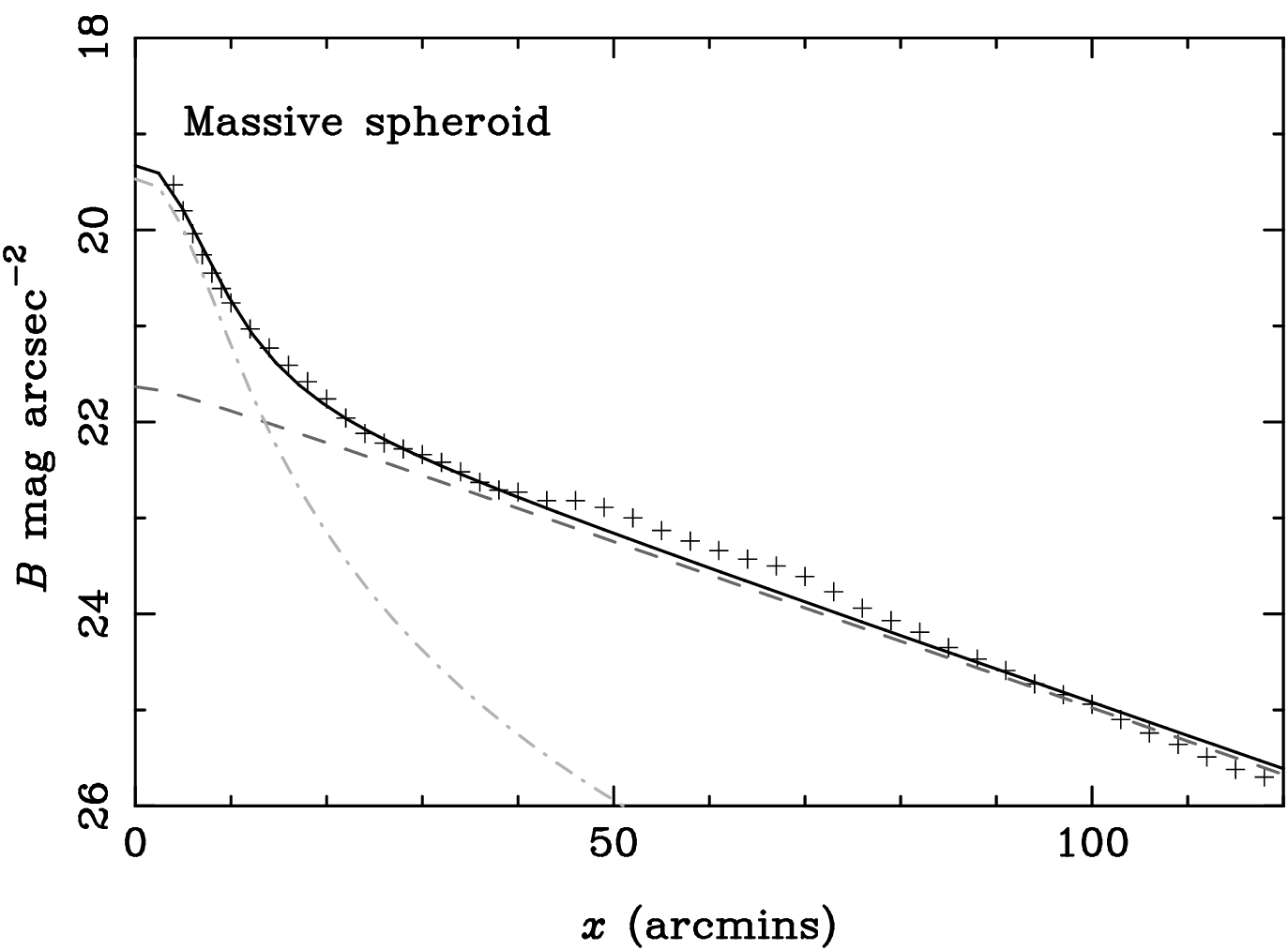}
\includegraphics[scale=0.3,angle=270]{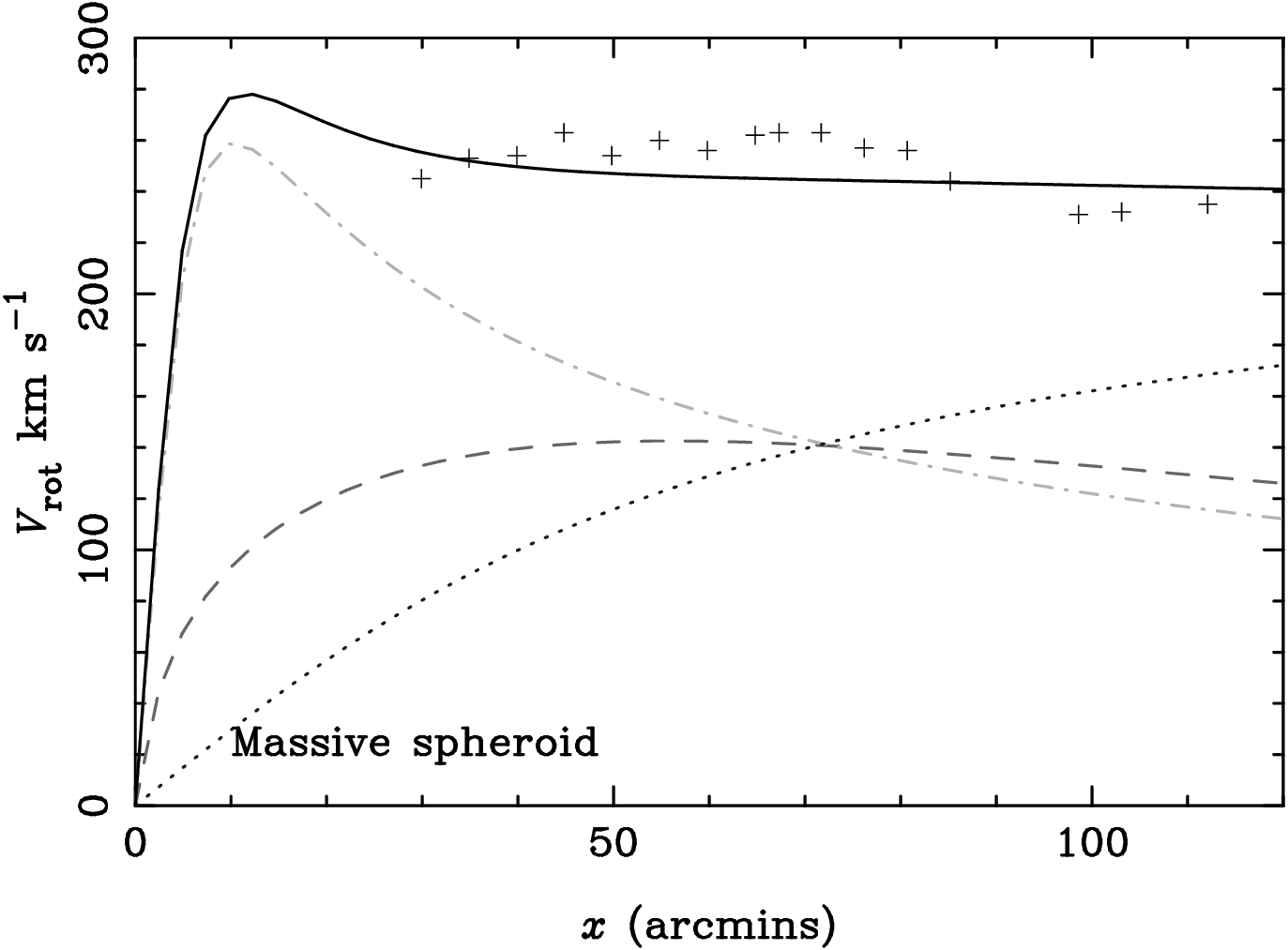}
\includegraphics[scale=0.3,angle=270]{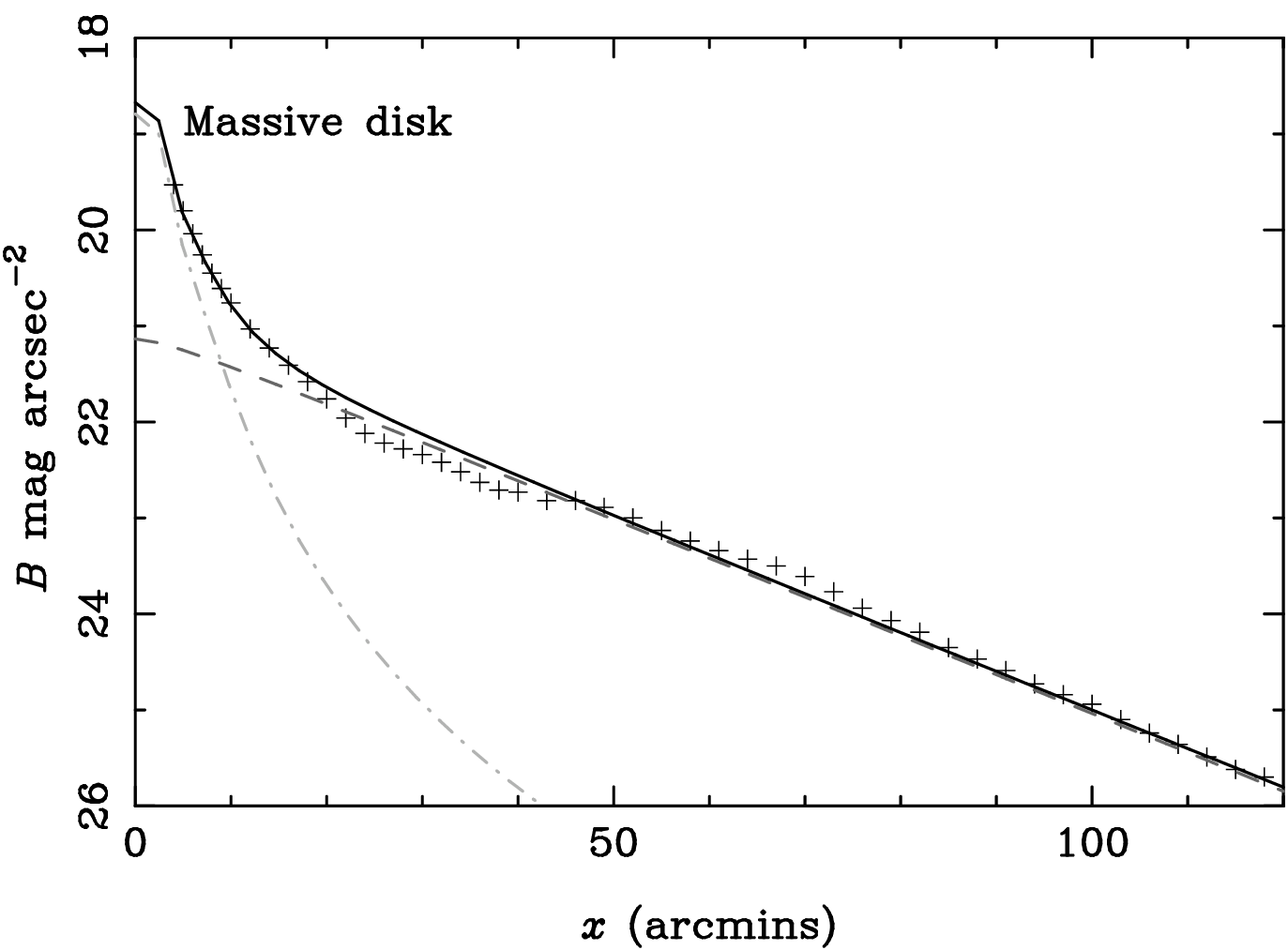}
\includegraphics[scale=0.3,angle=270]{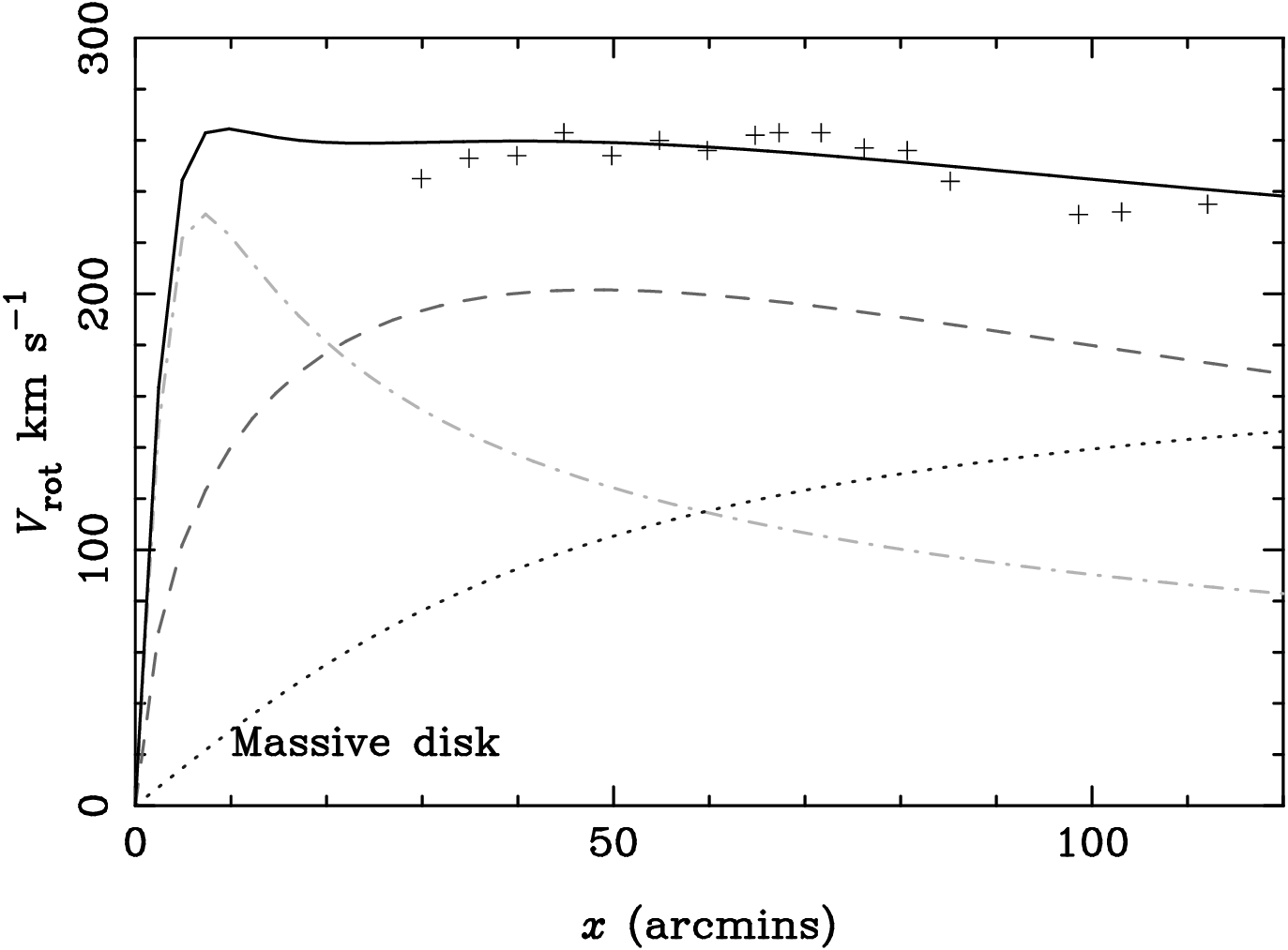}
\includegraphics[scale=0.3,angle=270]{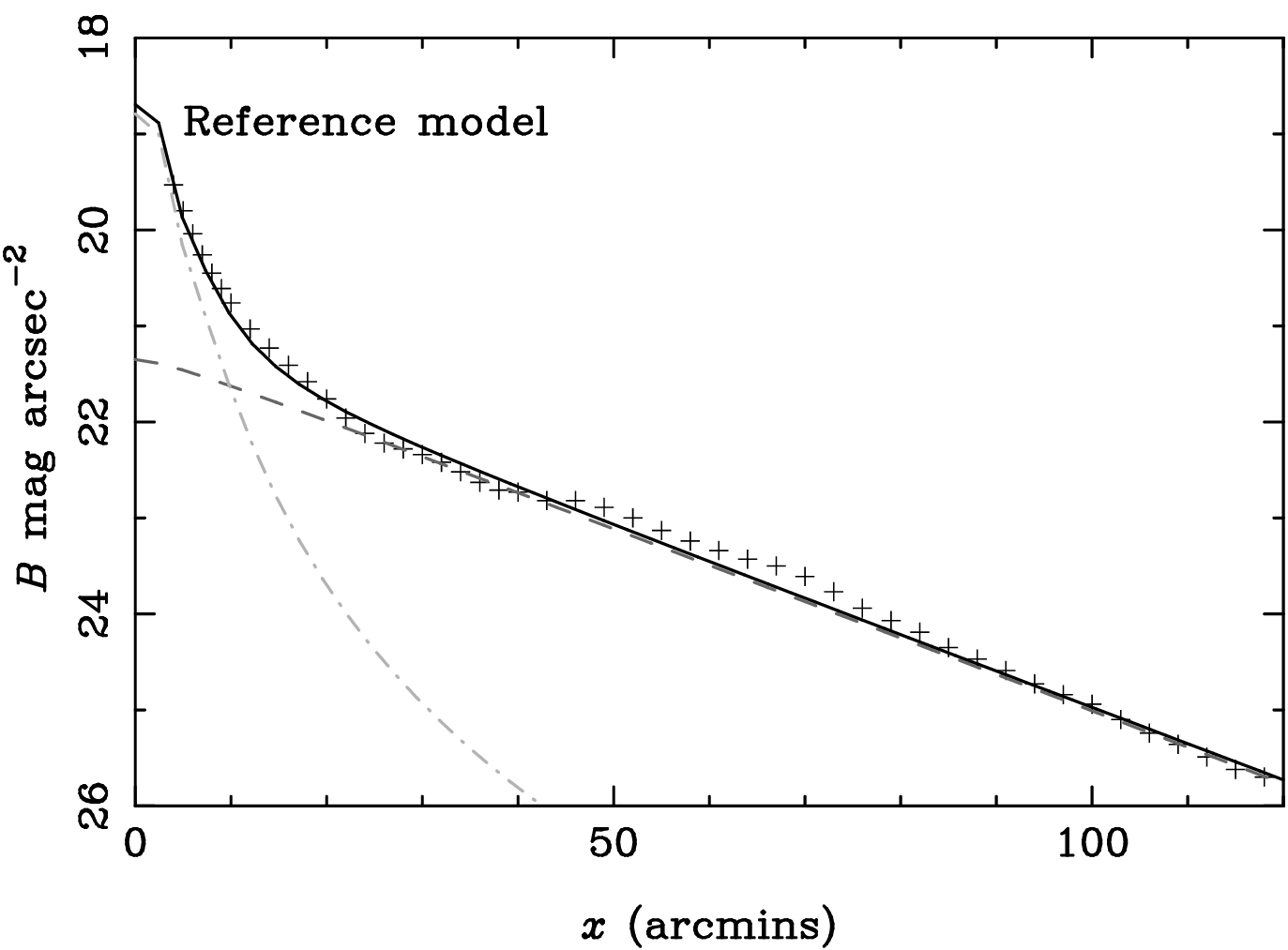}
\includegraphics[scale=0.3,angle=270]{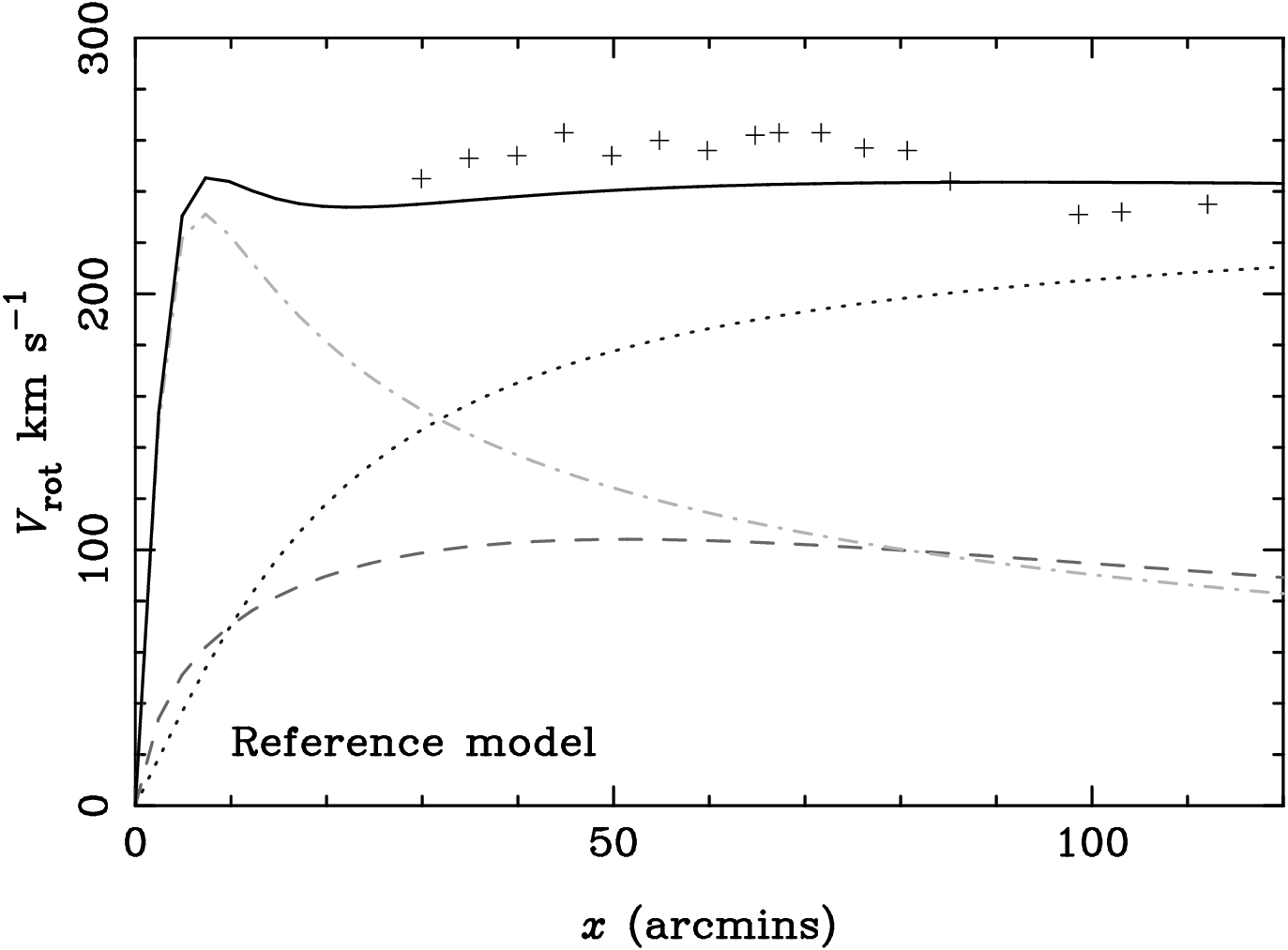}
\caption{ The $B$-band M31 major axis surface brightness profile
(left) and rotation curve (right) for the massive halo model,
massive spheroid model, massive disk model, and reference model. The
individual contributions of the spheroid, disk and halo are shown by
the light dot-dashed line, medium dashed line and dark dotted line,
respectively, whilst the overall profile is shown by the black solid
line.  The surface brightness measurements of \citet{wal87} and rotation curve
data of \citet{kent89} are shown by the crosses.}
\label{sb-rot}
\end{figure*}

The resulting surface brightness profiles and rotation curves of the
models are plotted in Figure~\ref{sb-rot}. The left-hand figures
show the predicted surface brightness distribution for the models,
including the individual contributions of the spheroid and disk. Also
plotted on these figures are the radially-averaged measurements of
\citet{wal87}. All models provide a remarkably good fit to the
data. The right-hand plots show the rotation curve contributions for
the spheroid, disk and halo, along with the overall profile.
Rotation curve data, taken from \citet{kent89}, are also
displayed. Whilst there are differences in the rotation curves of the
models, they provide reasonable fits to the data given
their simplicity.

We must emphasize that the assumed spheroid density law in equation~(\ref{spheroid}) is unlikely to provide an accurate description towards the inner galaxy, despite the good agreement with the surface brightness profile in 
Figure~(\ref{sb-rot}). However, this is not particularly a problem for our study of the distribution of microlensing events in the outer regions. \citet{rei98} found that the spheroid luminosity density in M31 is about ten times that of the Milky Way at a location equivalent to the Solar position in the Milky Way. For the spheroid model in the massive halo and disk models, which employs a standard spheroid $M/L_B$ value, we set the mass density normalisation to be consistent with this ratio. For the massive spheroid model the mass density is twice as large due to the larger assumed $M/L_B$. 

It should be stressed that the assumed form of the populations, especially that of the halo, plays a crucial role, along with the rotation curve and surface brightness distribution, in determining the relative masses of the components within each model. For example, in the massive halo model there is some leeway in the allowed ratio of spheroid to disk mass which would still produce acceptable profiles. Where this is the case we have tended to maximise the spheroid contribution. However, if we were to allow for dark halo distributions with density profiles which could be steeper or shallower than $r^{-2}$ then obviously an even greater range of spheroid and disk mass ratios would be permissible. Our conclusions must therefore be viewed within the context of our model assumptions.

\section{Microlensing event maps}\label{tau}

Whilst our three-component models satisfy the observed rotation curve
and surface brightness profile, we must remain aware of the
limitations of their simplicity. \citet{wid03} have undertaken a
detailed investigation of a number of self-consistent three-component
bulge-disk-halo models of M31 employing a range of additional
constraints including the inner velocity dispersion, bulge and disk
$M/L$, dynamical mass estimates, disk stability and cosmological
considerations. Such analyses are especially important in constraining
the expected rate of microlensing.  However, we can probe the relative
contributions of each population within our models by instead
considering the optical depth, $\tau$, a quantity which depends only
upon the density distribution of lenses and sources. In this sense it
is a more robust quantity than the event rate which depends
additionally upon the lens and source velocity distributions as well
as the lens mass function. Physically, $\tau$ is the number of ongoing
microlensing events per source star at any instant in time. So, we
can easily compute the instantaneous event number density as a
function of position by multiplying $\tau$ by the number density of sources.

\subsection{Pixel lensing event density and asymmetry} \label{denasymm}

Pixel lensing differs from conventional microlensing in that the
source stars are mostly unresolved, so microlensing events are seen
only as small enhancements of the local surface brightness flux. The
spatial distribution of pixel lensing events therefore depends upon
the source luminosity function and the background surface brightness
distribution, in addition to the usual parameters of classical
microlensing. Specifically, a sufficiently high event magnification
requires that the impact parameter, $u$, in units of the Einstein
radius be less than some threshold, $u_{\rm T}$, where $u_{\rm T}$
depends upon the unlensed source flux $F$ and the background surface
brightness at the event location. The number of microlensing events
which are detectable at some sky position $(x,y)$ will therefore
depend upon the quantity
   \be
	\overline{u}_{\rm T}(x,y) = \frac{\int u_{\rm T}(F,x,y)
	\phi(F) \, dF}{\int \phi(F) \, dF}, \label{impact}
   \ee
where $\phi$ is the source luminosity function. Since the optical
depth of events with impact parameter less than $u$ scales as
$u^2$, we define the pixel-lensing optical depth to be
   \be
	\tau_{\rm p} = \overline{u}_{\rm T}^2 \tau.
   \ee
This quantity is the number of ongoing events with $u <
u_{\rm T}$ at a given location.

We compute $\tau_{\rm p}$ over a grid of sky positions $(x,y)$ on the
M31 disk for each of the models in Table~\ref{tab-param}, and for each
combination of lens and source population. The luminosity function of
M31 stars is known only for the brightest stars so, to calculate
$u_{\rm T}$, we make $V$-band calculations taking M31 sources to have
a Solar neighbourhood luminosity function for $M_V > 0$
(\citealt{bah80,wie87}) and use the M31 luminosity function determined
by \citet{hol96} for $M_V \leq 0$. Whether a microlensing-induced flux
enhancement is detectable depends upon experimental sensitivity, so
$\tau_{\rm p}$ is implicitly an experiment-specific quantity. We base
our computations loosely on the POINT-AGAPE survey \citep{aur01} by
computing flux changes within a square pixel array (a ``super-pixel'')
with a $2\farcs3$ field of view. The radially-averaged surface
brightness measurements of \citet{wal87} are used to compute the M31
surface brightness within the super-pixel. We assume that typically
$40\%$ of a randomly-positioned point source PSF is contained within
the super-pixel and that the survey is sensitive to super-pixel flux
variations exceeding $1\%$. In addition to the M31 surface brightness,
we assume a sky background in $V$ of 19.5~mag/arcsec$^2$, which
conservatively allows for a lunar sky background contribution.

\begin{table*}
\begin{minipage}{160mm}
\caption{The number of ongoing events and the number asymmetry for the
three extreme M31 models and for the reference model. The asymmetry is expressed both in absolute numbers of events and as a ratio (in brackets). The figures assume a full Macho halo and
apply to the central $100 \times 70$~arcmin$^2$ region orientated
along the major axis of M31 (see Figure~\ref{maxmodels}). We exclude
events occurring within a radius of 8~arcmin of the M31 centre.}
\label{tab-asymm}
\begin{tabular}{@{}lccccccccc}
\hline
 & \multicolumn{4}{c}{No. of events ($100\times 70$~arcmin$^2$)} & &
   \multicolumn{4}{c}{Asymmetry (far disk/near disk)} \\
Model & Halo & Spheroid & Disk & Overall & & Stellar & Halo & Halo-to-disk
 & Overall \\
\hline
Massive halo & 1.27 & 0.12 & 0.07 & 1.46 & & 0.12/0.07 (1.67) & 1.01/0.26 (3.92) & 0.83/0.16 (5.24) & 1.13/0.33 (3.43)\\
Massive spheroid & 0.70 & 0.30 & 0.27 & 1.27 & & 0.29/0.28 (1.02) & 0.51/0.19 (2.75) & 0.37/0.09 (3.87) & 0.80/0.47 (1.72) \\
Massive disk & 0.51 & 0.12 & 0.41 & 1.04 & & 0.21/0.32 (0.68) & 0.38/0.13 (3.07) & 0.32/0.08 (3.77) & 0.60/0.44 (1.36) \\
\hline
Reference & 1.08 & 0.11 & 0.11 & 1.31 & & 0.12/0.10 (1.22) & 0.85/0.23 (3.60) & 0.69/0.14 (4.80) & 0.97/0.34 (2.89) \\
\hline
\end{tabular}
\end{minipage}
\end{table*}

Table~\ref{tab-asymm} displays the total number of ongoing events over
a central $100 \times 70$~arcmin$^2$ region of M31, within which
current surveys are looking for Machos. We have assumed for this Table
that the M31 halo is full of Machos; for a partial Macho halo the
number of halo events scales linearly with halo fraction. Since we are
interested in the effect of stellar lenses on the spatially extended
Macho population we have defined an 8~arcmin radius exclusion zone,
counting only events outside this region. Since the inner region may
be dominated by stellar lensing, surveys are likely to define such
masks as a simple way of minimising the impact of the stellar lens
contribution, so making their results less sensitive to the precise
details of the inner galactic structure. The left-hand columns of the
Table shows the number of ongoing events for each population, together
with the cumulative total. The massive halo model provides the largest
number of events overall, thanks to the substantially more massive
halo assumed for this model. For all three extreme models M31 Machos provide
the largest contribution of any individual lens population. However,
whilst M31 Machos contribute almost $90\%$ of the event density for the
massive halo model, they provide only around half of the events for
the massive spheroid and disk models.  In the far disk, beyond the
region considered, the relative contribution of M31 Machos is
significantly larger but the number of available sources is few and
these regions are not being targeted by current surveys.

The disk shows a factor of 6 variation in the expected number
of events between the three extreme models, whilst for the halo and spheroid
the variation is just over a factor of 2. The variation in the overall
number of events is rather less, at around $40\%$. As we shall discuss
in Section~\ref{rate}, the number of events computed from $\tau_{\rm
p}$ does not necessarily reflect the relative contributions one might
observe in the event rate, since the rate also depends upon the event
durations which are sensitive to the lens and source velocity
distributions, as well as the lens mass function. Differences in these
distributions between the lens populations will therefore mean
differences between the relative contribution to the rate and the
contribution to $\tau_{\rm p}$.

The right-hand columns of Table~\ref{tab-asymm} show the asymmetry in
the number of ongoing events against the far disk to those against the
near disk. The numbers for the halo population again assume a halo
full of Machos. The signature of asymmetry is the key diagnostic for
the identification of Machos, but as Table~\ref{tab-asymm} shows, one
may also see an asymmetry in spatial distribution of stellar
events. In the massive halo model the ratio of far-disk to near-disk
events is 3.9 for M31 Machos and 1.7 for stellar lenses. Stellar
lenses in this model therefore have a spatial asymmetry which is
weaker but in the same direction as that of M31 Machos. For the
massive disk model the stellar asymmetry is reversed, with only $40\%$
of stellar events occurring on the far side of the disk. This is
because the disk in this model has a very high $M/L$ and so disk
lensing against spheroid sources dominates the number of stellar
events. In the massive spheroid model the $M/L$ of the spheroid and
disk components are comparable and the distribution of stellar events
is close to symmetric. These results bear out the analytic prediction
of Section~\ref{asymm} that the presence and direction of spatial
asymmetry due to stellar lenses is dictated by the ratio of $M/L$
values in the spheroid and disk populations. Clearly, unless there is
fine tuning in the spheroid and disk $M/L$, one should expect some
asymmetry in the spatial distribution of stellar lenses. However, this
asymmetry should be weaker than that due to M31 Machos if Machos
contribute significantly to the dark matter.

The column labelled ``Halo-to-disk'' in Table~\ref{tab-asymm} shows
the asymmetry of M31 Machos against disk sources alone, ignoring the
effect of spheroid sources. Comparison with the adjacent column labelled
``Halo'', which includes both disk and spheroid sources, indicates
that the presence of the spheroid sources dilutes the M31 Macho
asymmetry by $20-30\%$ for the three extreme models. This result is largely
insensitive to the very different mass normalisations assumed for the
three extreme models.

\begin{figure*}
\includegraphics[scale=0.3,angle=270]{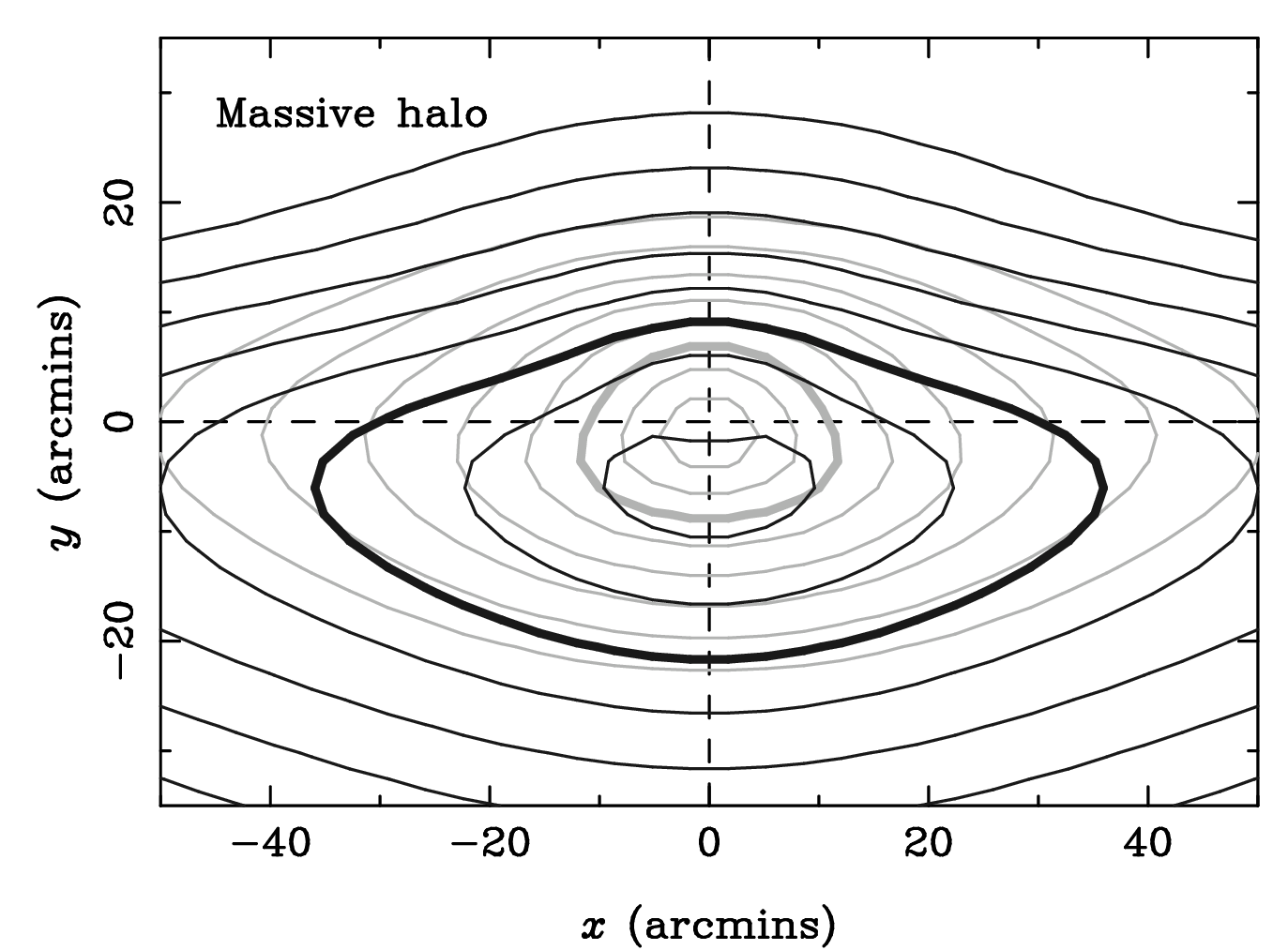}
\includegraphics[scale=0.3,angle=270]{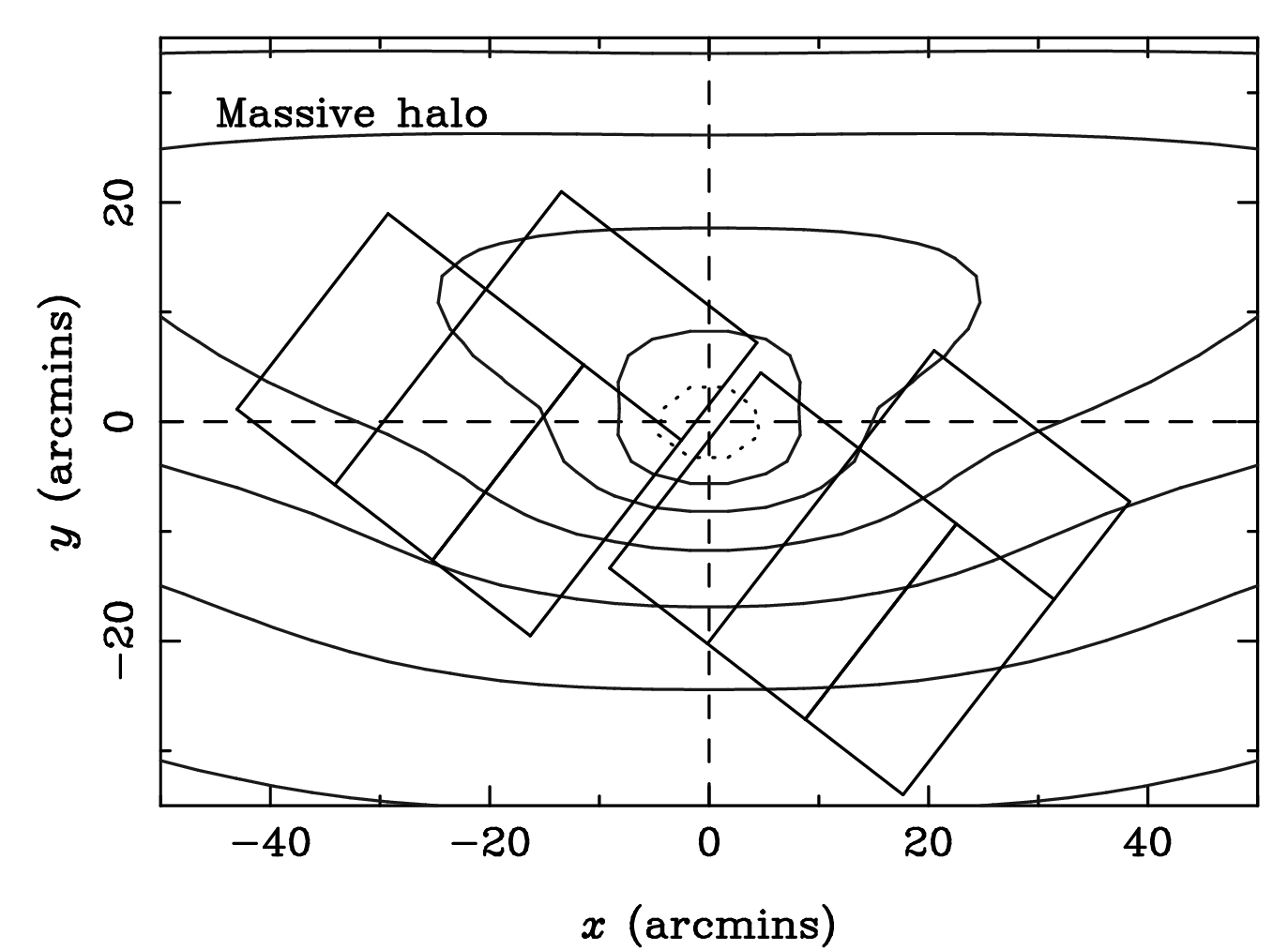}
\includegraphics[scale=0.3,angle=270]{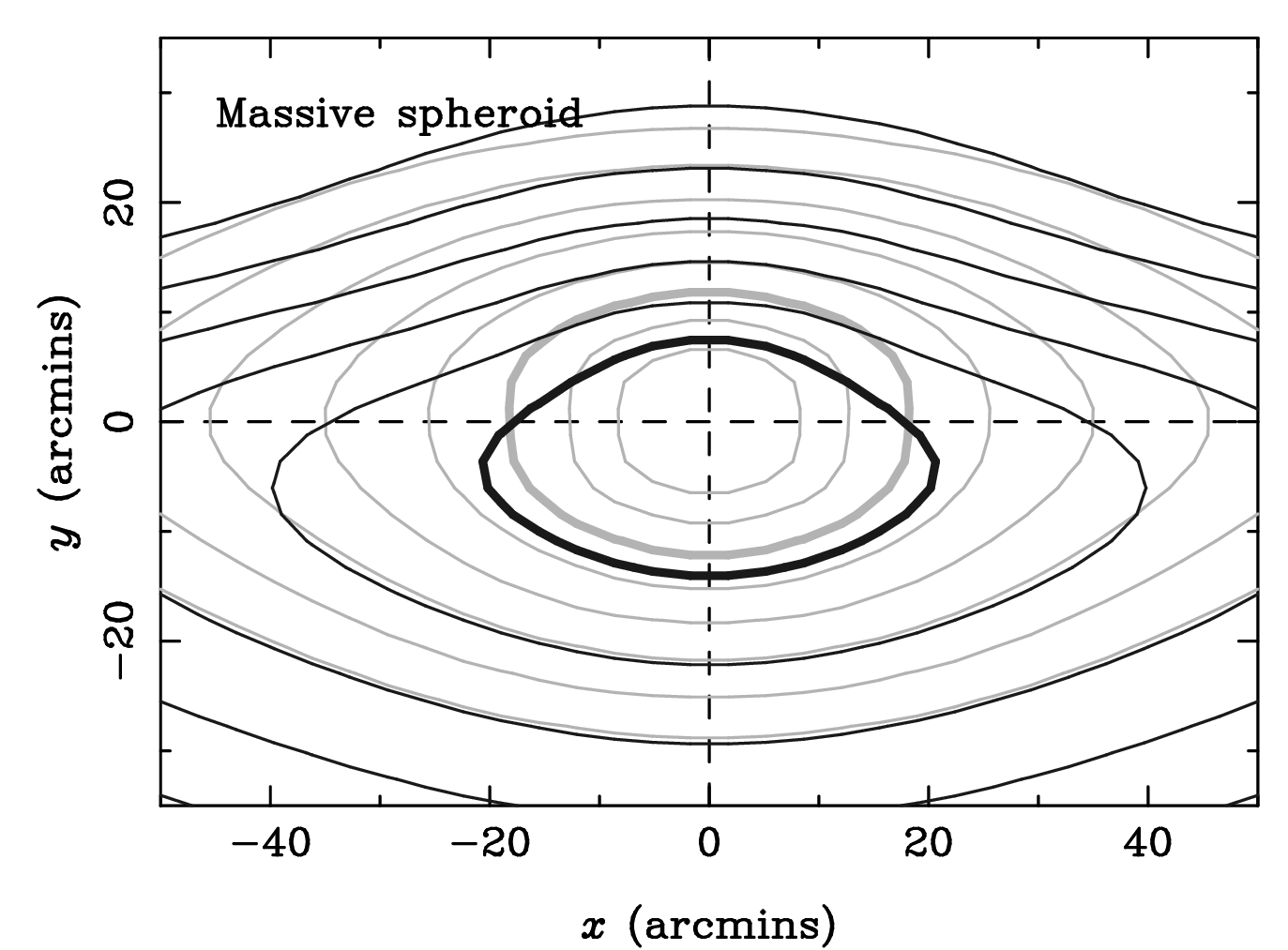}
\includegraphics[scale=0.3,angle=270]{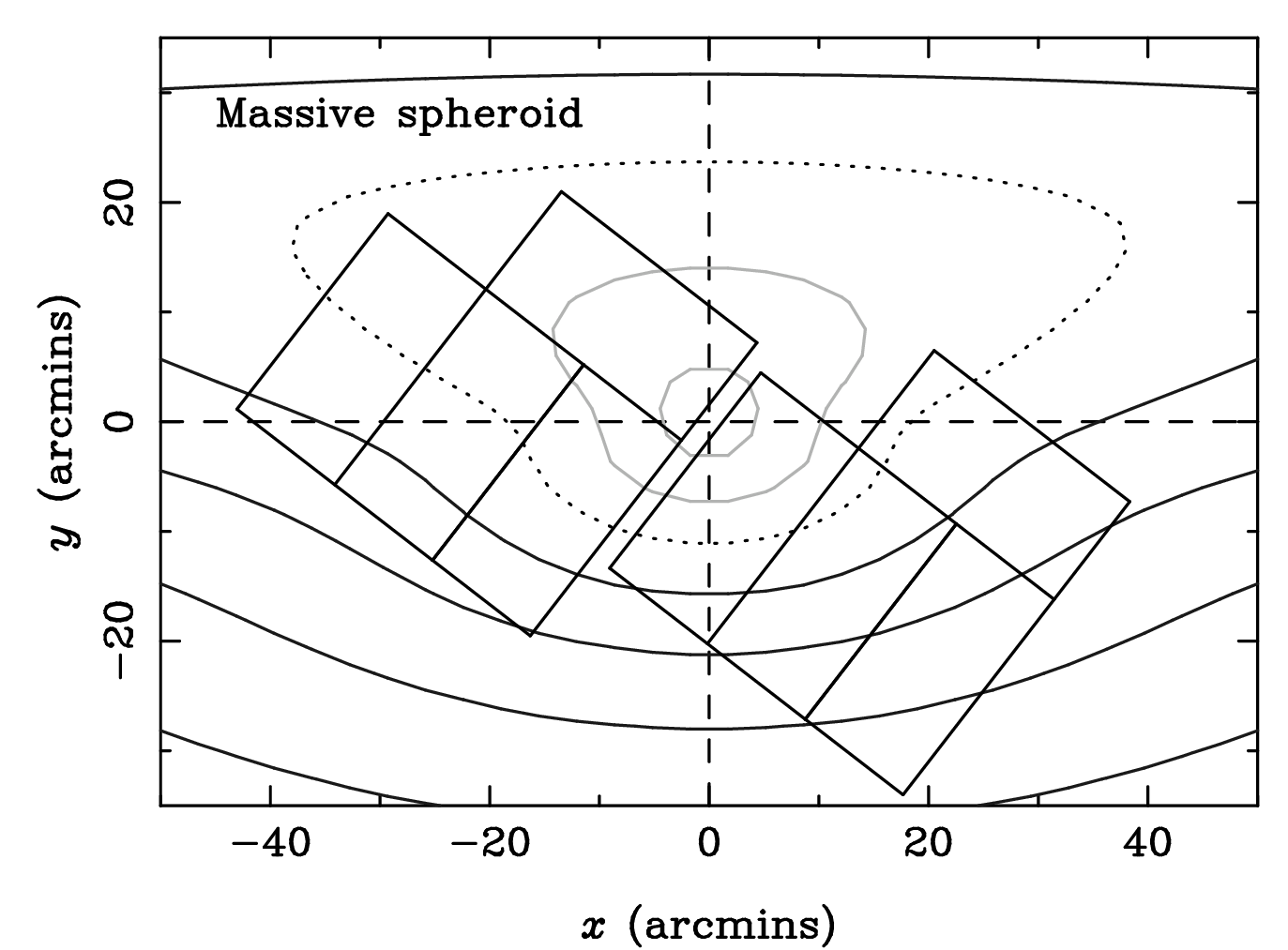}
\includegraphics[scale=0.3,angle=270]{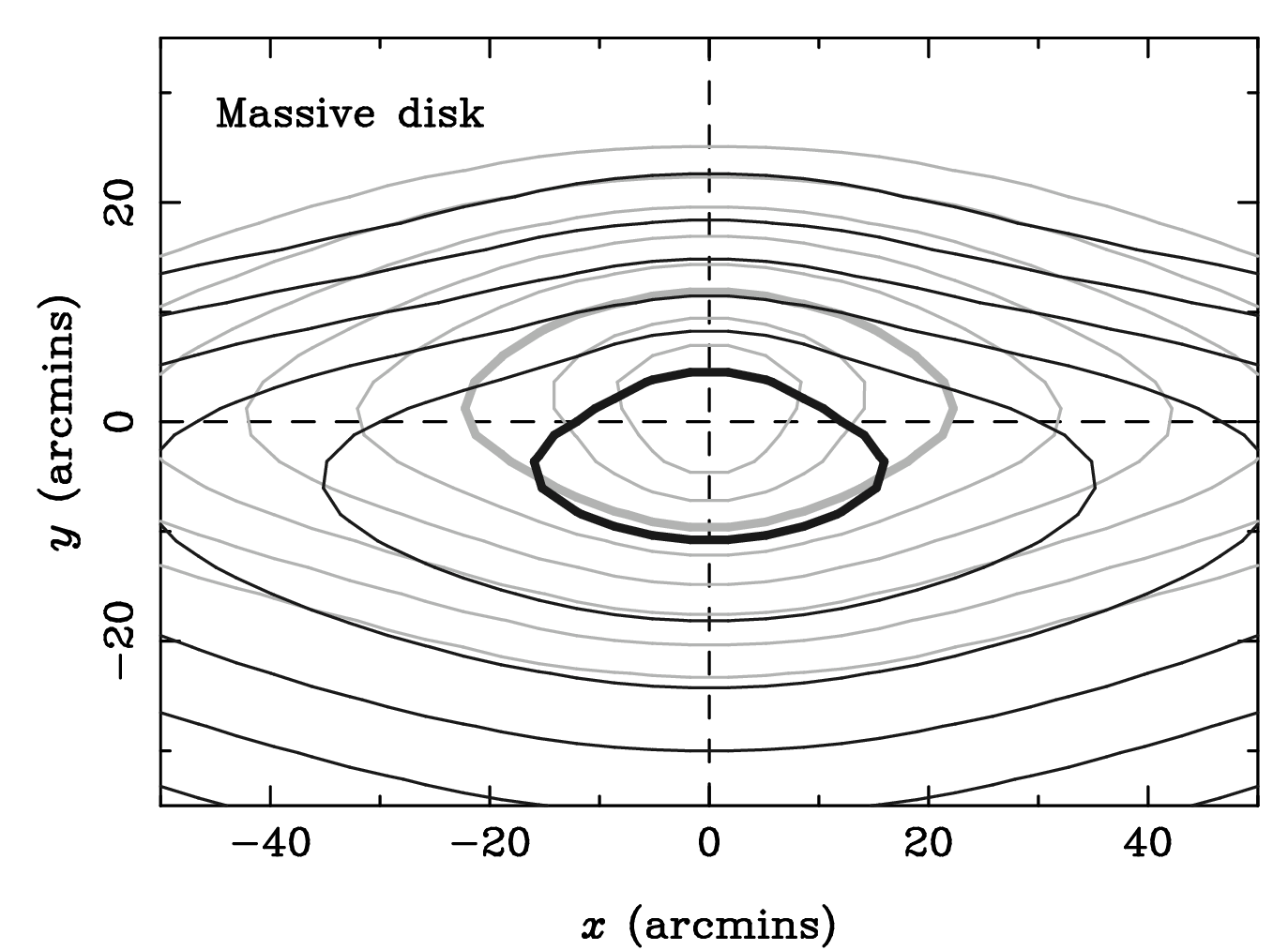}
\includegraphics[scale=0.3,angle=270]{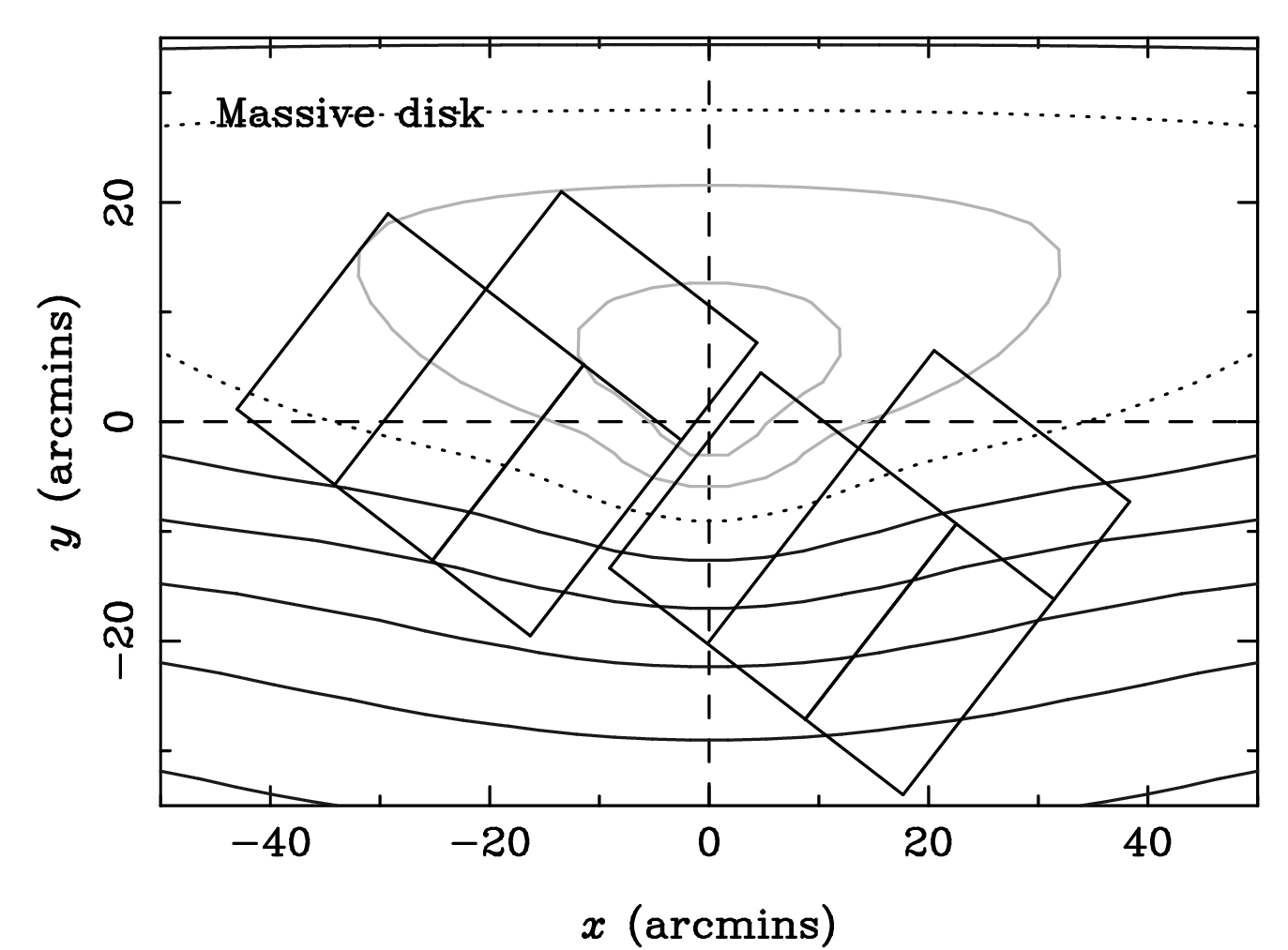}
\includegraphics[scale=0.3,angle=270]{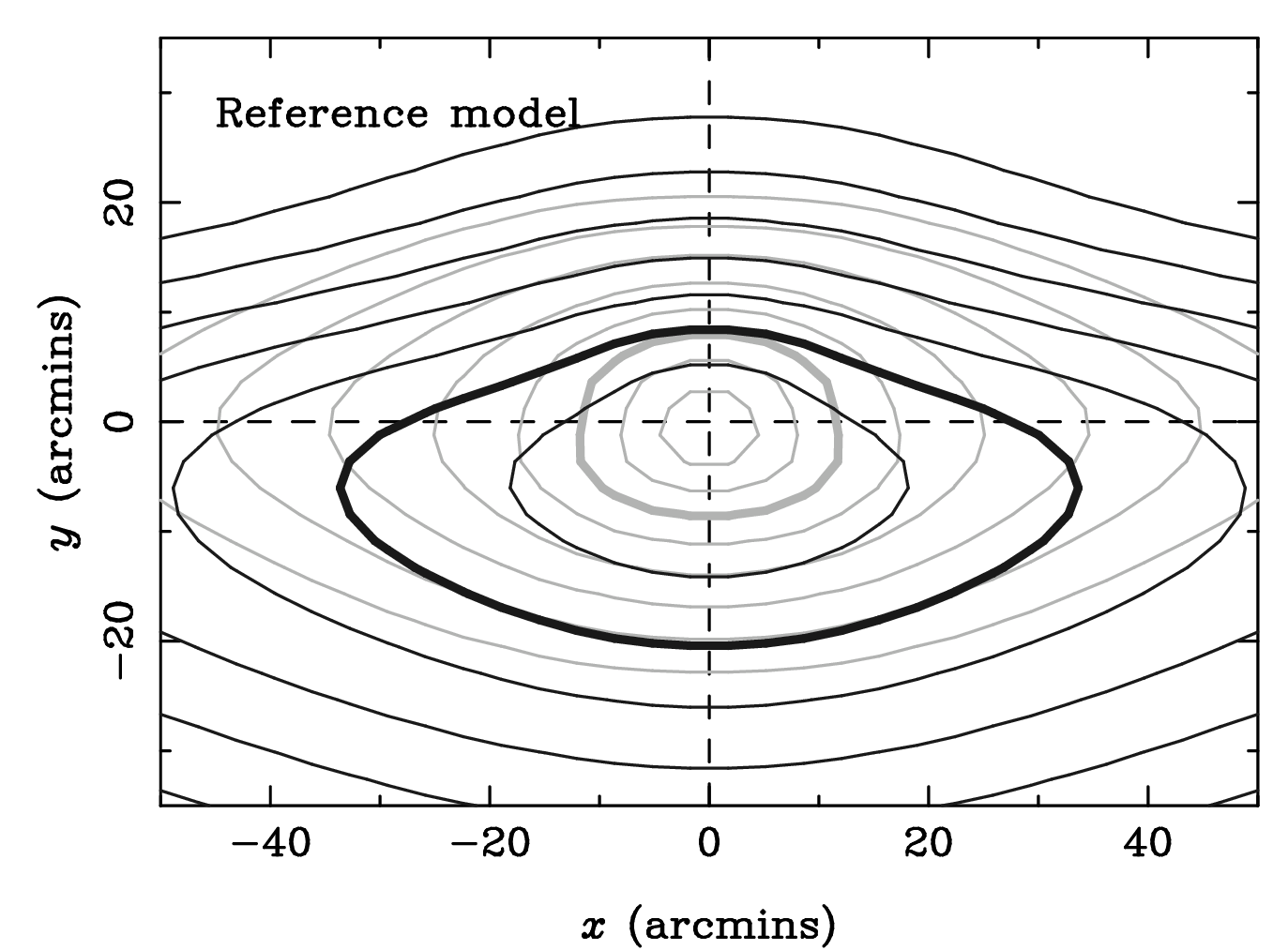}
\includegraphics[scale=0.3,angle=270]{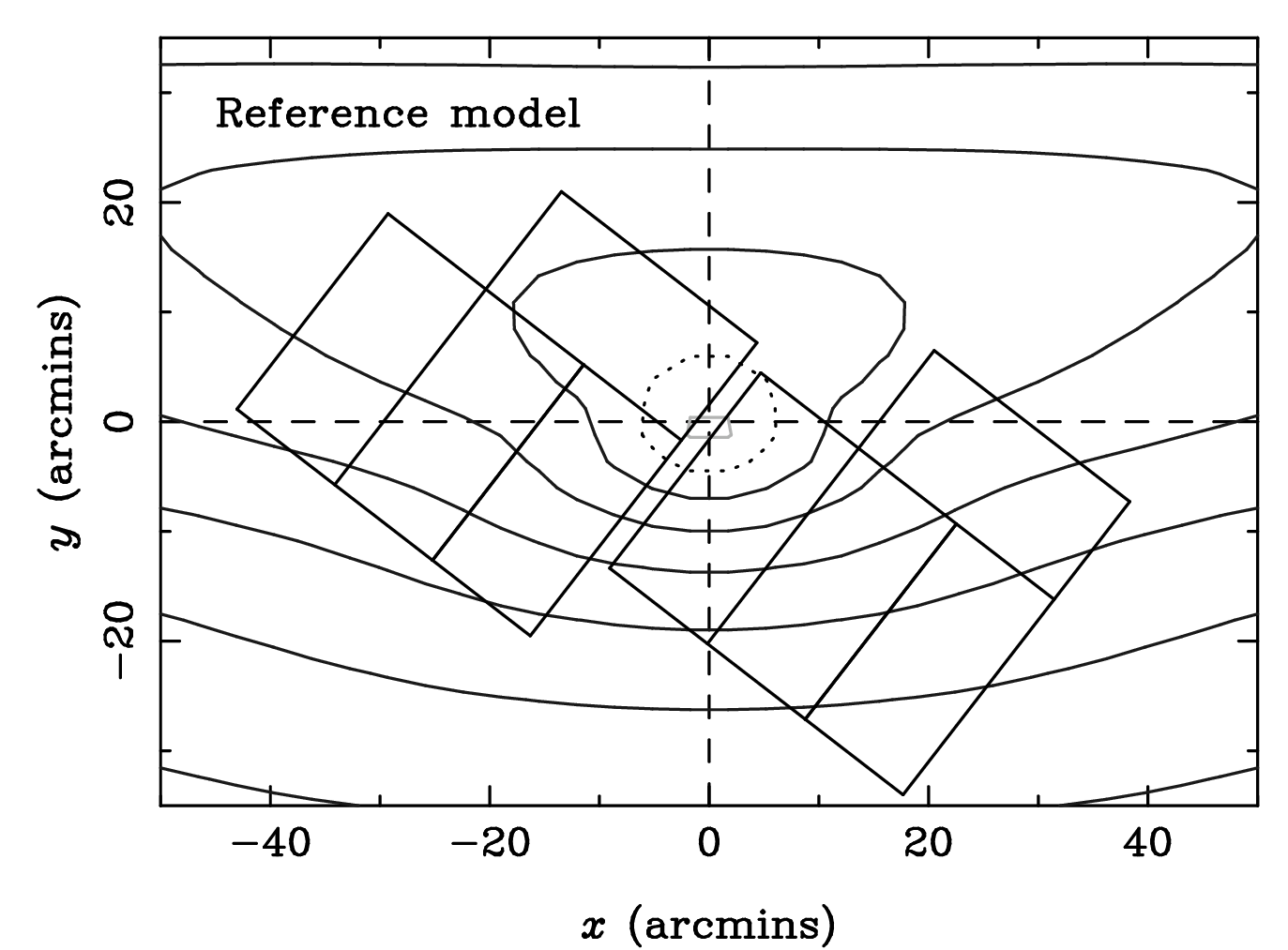}
\caption{
M31 contour maps of instantaneous pixel-lensing event number density
(left) and of relative Macho-to-stellar event number density (right)
for the massive halo, massive spheroid, massive disk
and reference models, assuming a full Macho halo. The total event number density
contribution due to M31 Macho lenses is shown by the dark contours in
the left-hand panels, whilst the lighter contours show the contribution
due to stellar lenses. The thick contours represent a density of
$3\times 10^{-4}$~events/arcmin$^2$. Successive contours represent a
factor 2 change in event density.  The ratio of the Macho and stellar
contributions is shown in the right-hand panels. The dark contours
show regions where M31 Macho lenses dominate, whilst lighter contours
show where stellar lensing dominates. The black dotted contour shows
where the M31 Macho and stellar lens contributions are equal. Again,
successive contours represent a factor 2 ratio change. The field
positions of the POINT-AGAPE/MEGA INT survey are also shown.
}
\label{maxmodels}
\end{figure*}

In Figure~\ref{maxmodels} the left-hand panels display contours of
event number density, that is $\tau_{\rm p}$ multiplied by source
number density, for both M31 Macho and stellar lensing events over a
central area covering $100\times 70$~arcmin$^2$. A full Macho halo is
assumed. The $x$ and $y$ axes are orientated along the major and minor
axes of M31, respectively, with positive $y$ pointing towards the near
side of the disk. The dark contours show the event density of M31
Machos, with the thick contour indicating a density of $3 \times
10^{-4}$~events/arcmin$^2$. The density increases towards the M31
centre by a factor of 2 for each successive contour. The near-far
asymmetry signature of the M31 Macho population is strongest for the
massive halo model (top) but is clear for all three extreme models.  The
different mass normalisations of each component in the
three extreme models have a clear impact on the number of M31 Machos, as
evidenced by the relative positions of the thick contours.

The stellar lens distribution is shown by the light grey contours in the
left-hand panels of Figure~\ref{maxmodels}. Again, the thick contour
indicates an event density of $3 \times 10^{-4}$~events/arcmin$^2$ and
contour spacing is the same as for the Macho distribution. As
expected, the contours for the massive disk model are substantially
flattened compared to those for the massive spheroid model. The
stellar asymmetry for the massive halo and massive disk models is
clearly evident. In the massive halo model the M31 Macho distribution
is clearly more extended than that of the stellar lenses, however for
the massive spheroid and disk models the stellar contribution remains
substantial out to large distances from the M31 major axis.

The relative importance of the stellar and Macho contributions to the
number of ongoing events varies considerably across the three extreme models,
as indicated by the right-hand panels of Figure~\ref{maxmodels}. The
light grey contours show the region where stellar lensing dominates, whilst
darker contours indicate areas where M31 Machos provide the dominant
contribution, assuming a full Macho halo. The locus of positions where
the stellar and M31 Macho event number densities are equal is shown by
the dotted line. Inside the dotted line the ratio of stellar-to-Macho event densities doubles as one progresses towards the M31 centre, whilst outside the dotted line
the stellar-to-Macho event ratio halves for each contour level going outwards.  The
positions of the INT fields surveyed by the POINT-AGAPE and MEGA teams
are also indicated for comparison (\citealt{aur01,cro01}). In the
inner regions the event density is dominated by M31 Machos only for
the massive halo model (top). In the massive spheroid model (middle
panel) stellar lenses dominate within the central $10-15$~arcmins and
also over a large area of the near disk between $|x| \la 30$~arcmins
and $0 \la y \la 20$~arcmins. For the massive disk model
stellar events provide the bulk of the near-disk events over almost
the whole region considered. However, even for these extreme models
Machos dominate the far disk event numbers for the case of a full
Macho halo. For a halo with only a $50\%$ halo fraction the dotted
contour is shifted outward by one contour level and for a $25\%$ halo
fraction it is shifted outward by two levels.

From this we can estimate how small the M31 Macho halo fraction would
have to be for the POINT-AGAPE/MEGA INT fields to be dominated
everywhere by stellar rather than Macho lensing events. For the
massive halo model Machos should dominate at least part of these
survey fields provided their halo fraction exceeds $2\%$. For the
massive disk model we require a halo fraction above $4\%$ and for the
massive spheroid model the Macho fraction must be at least $8\%$. More
generally, at least half of the far-disk region considered must be
dominated by Machos if the M31 Macho halo fraction exceeds about
$25\%$.  Interestingly, though stellar lenses dominate the largest
fraction of the near disk in the massive disk model, their density
drops off faster in the far disk than in the massive spheroid model,
which is why the spheroid model requires a larger Macho halo fraction.

The event density distribution for the reference model is also shown in Figure~\ref{maxmodels}. The distribution for this model is intermediate to those of the massive halo and spheroid models. The event density is dominated by Machos for this model outside of the innner 5 arcmins, assuming a full Macho halo, and dominates both the near and far disk. In this respect, the massive disk model may be regarded as the {\em most} extreme and unlikely of our three extreme models.

\subsection{Event rates and timescales in pixel lensing} \label{rate}

The pixel-lensing optical depth, $\tau_{\rm p}$, can be related to
pixel-lensing timescales and rates in much the same way as the
analogous observables in classical microlensing. We can define the
pixel-lensing rate as $\Gamma_{\rm p} = \overline{u}_{\rm T} \Gamma$
\citep{ker01}, where $\Gamma$ is the classical microlensing
rate. Whilst in classical microlensing the characteristic event
duration is given by the Einstein Radius crossing time, $\te$, in
pixel lensing $\te$ is generally not measurable because of a
near-degeneracy in the light-curve between the event timescale,
magnification and source luminosity \citep{gou96}. Instead the usual timescale
observable in pixel lensing is the full-width half-maximum duration of
the event. This can attain a maximum given by
   \begin{eqnarray}
	t_{1/2}^{\max} & \simeq & 2\sqrt{3} u_{\rm T} \te \quad \quad
	(u_{\rm T} \ll 1) \nonumber \\ & \simeq & 17~\mbox{days}~
	\left( \frac{u_{\rm T}}{0.1} \right)
	\left(\frac{m}{\sm}\right)^{1/2} \left( \frac{D}{\mbox{kpc}}
	\right)^{1/2} \nonumber \\ & & \quad \quad \quad \quad \quad
	\quad \quad \quad \quad \times \left( \frac{v}{100~\mbox{km
	s}^{-1}} \right)^{-1}, \label{pixtime}
   \end{eqnarray}
where $m$ is the lens mass, $v$ is the lens transverse velocity
relative to the line of sight and $D \simeq \ds - \dl$. With these
definitions we have
	\be
	   \tau_{\rm p} \simeq \frac{\pi}{2\sqrt{3}}
	   \overline{t}_{1/2} \Gamma_{\rm p}, \label{pixeq}
	\ee
where $\overline{t}_{1/2} \simeq 2 \sqrt{3} \overline{u}
\overline{t}_{\rm E}$ is the average $t_{1/2}$ timescale and over-lines
on other quantities also denote averages. Equation~(\ref{pixeq})
relies on the fact that $\overline{u} = u_{\rm T}/2$, so $t_{1/2} =
t_{1/2}^{\max}/2$ on average for fixed $\te$.

\begin{table*}
\begin{minipage}{105mm}
\caption{The estimated event rate, assuming $100\%$ detection
efficiency, for the three extreme M31 models. The figures assume a
full Macho halo and apply to the central $100 \times 70$~arcmin$^2$
region of M31, though excluding events within a radius of 8~arcmin of
the M31 centre . $D$, $v$ and $t_{1/2}$ are indicative values,
$N(\tau_{\rm p})$ is the number of ongoing events computed from
$\tau_{\rm p}$ and $N(\Gamma_{\rm p})$ is the estimated event rate,
taking the indicative $t_{1/2}$ to be an estimate of
$\overline{t}_{1/2}$. A lens mass of $0.5\,\sm$ is assumed for
$t_{1/2}$ and $\Gamma_{\rm p}$. The numbers in brackets correspond to
predictions for the massive (halo, spheroid, disk) models,
respectively.}
\label{tab-rate}
\begin{tabular}{@{}lccccc}
\hline
Lens/Source & $D$/kpc & $v$/km~s$^{-1}$ & $t_{1/2}$/days &
 $N(\tau_{\rm p})$ & $N(\Gamma_{\rm p})$/year \\
\hline
halo/disk & 10 & 300 & 6 & (1.24,0.56,0.50) & (63,29,25) \\
halo/spheroid & 10 & 290 & 7 & (0.28,0.25,0.11) & (17,15,7) \\
spheroid/disk & 1 & 240 & 2 & (0.12,0.25,0.12) & (15,29,15) \\
spheroid/spheroid & 2 & 240 & 4 & (0.03,0.12,0.03) & (3,14,3) \\
disk/disk & 1 & 60 & 10 & (0.03,0.08,0.19) & (1,3,8) \\
disk/spheroid & 1 & 240 & 2 & (0.05,0.26,0.30) & (7,30,36) \\
TOTAL & -- & -- & -- & (1.75,1.52,1.25) & (106,121,94) \\
\hline
\end{tabular}
\end{minipage}
\end{table*}

We can estimate the typical $t_{1/2}$ duration of pixel-lensing events
from very general considerations of the various lens and source
components. Though details of the lens and source velocity
distributions can vary even for distribution functions with similar
density profiles, the event duration represents a line-of-sight
statistical average, and so is a reasonably robust
quantity. Furthermore, the $t_{1/2}$ duration depends not just on the
lens and source distribution functions but also on the lens mass
function and source luminosity function. The $t_{1/2}$ timescale
distribution is therefore rather broad and variations in the lens or
source velocity distribution make a relatively modest contribution to
this width.

The underlying $\te$ depends upon the lens mass $m$, relative
transverse velocity $v$ and lens--source separation $D$. Additionally,
$t_{1/2}$ depends upon the impact parameter $u < u_{\rm T}$, where
$u_{\rm T}$ depends upon the source luminosity and background surface
brightness. For our computations of $\tau_{\rm p}$ in the previous
subsection, $u_{\rm T} \sim 0.1$ over most of the area considered. On
average we should find $u = u_{\rm T}/2 \sim 0.05$, so we adopt this
value for all $t_{1/2}$ estimates. The typical mass of stellar lenses
depends on their mass function $\phi(m)$. In the Solar neighbourhood
$\phi \propto m^{-2.35}$ for $m \ga 0.5\,\sm$ and $m^{-1.4}$ below
$0.5\,\sm$ \citep{gou97}. The lensing rate per unit lens mass is
$d\Gamma/dm \propto m^{1/2}\phi(m)$, and therefore the mean lens mass
for stars between $0.01\,\sm$ and $10\,\sm$ is $\overline{m}
\sim 0.5\,\sm$ in the Solar neighbourhood. In the absence of detailed
knowledge of the M31 stellar mass function we adopt $m = 0.5\,\sm$ for
M31 stars. This also corresponds to the mass scale for Machos favoured
by the results of the MACHO microlensing survey of the LMC
\citep{alc00}.

The relative transverse velocity $v$ varies according to lens and
source population. If lens and source velocities can be described by a
combination of rotational and random motions then we can expect the
characteristic velocity to be $v \sim [(v_{\rm sys,l}-v_{\rm
sys,s})^2+v_{\rm ran,l}^2+v_{\rm ran,s}^2]^{1/2}$, where subscript
``l'' and ``s'' denote lens and source velocities, respectively,
``sys'' denotes systematic motions due to rotation and ``ran''
represents the random velocity component. Table~\ref{tab-rate} lists
characteristic values for $v$ for each combination of lens and source
population. The numbers assume that Machos have 2-D random velocities
of 235~km~s$^{-1}$, comparable to the disk rotation speed. We also
assume disk and spheroid 2-D random velocities of 45 and
170~km~s$^{-1}$, respectively, and a spheroid rotational component of
30~km~s$^{-1}$. The contribution of rotational motions to $v$ will
depend upon where events occur against the inclined M31 disk but, on
average over the disk, the transverse component is about $80\%$ of
the total rotational motion.  Table~\ref{tab-rate} also indicates
typical values for the lens--source separation $D$.

Estimates for the typical $t_{1/2}$ and $\Gamma_{\rm p}$ are computed
from these representative values of $m$, $v$ and $D$, together with an
assumed impact parameter $u = 0.05$. Based on these numbers, for disk
lenses we can expect $N \sim 8$~events/yr for the massive halo model,
or up to six times this number in the case of the massive disk model.
For spheroid lenses we can expect $\sim 18$~events/yr from the massive
halo and disk models ranging up to 43~events/yr for the massive
spheroid model. The expected number of M31 Machos ranges from $\sim
32$~events/yr for the massive disk model up to $\sim 80$~events/yr for
the massive halo model.

The ratio of halo:spheroid:disk event rates is
1:0.2:0.1 for the massive halo model, 1:1:0.75 for the massive
spheroid model and 1:0.6:1.4 for the massive disk model. These numbers
indicate that in the massive disk and spheroid models stellar events
are expected to dominate the overall rate. This is because the disk
and spheroid in these models are effectively repositories of large
amounts of dark matter in their own right. For the massive halo model,
which employs more conventional values for the disk and spheroid $M/L$,
Machos out-number stellar events by more than 3:1, with Machos becoming
increasingly dominant in the far disk. 

An important caveat to these estimates is that they assume a $100\%$
detection efficiency for events of all timescales, something which
is not the case for any real experiment. Sampling efficiency in
particular is likely to have a large impact on the detected number of
stellar events, which for the most part have typical durations of just
a few days. Current experiments are unlikely to have good detection
efficiency for events with durations below $t_{1/2} \sim
3$~days. Given that the $\log( t_{1/2})$ distribution is expected to be very
broad, and the fact that we estimate $\overline{t}_{1/2} \sim 3$~days
for most stellar events, we can expect that many more stellar events
will escape detection, as compared to Machos with $m \sim
0.5\,\sm$. This may mean that, even for the massive disk and spheroid
models above, Machos still dominate the rate of {\em detected}\/ events.
Additionally, finite source size effects are not taken into account in
these estimates, but are likely to be more important for stellar
lenses than for Machos, due to the smaller lens--source
separation. When $D$ is sufficiently small that the Einstein radius
has a comparable angular size to the source star one expects
differential magnification across the face of the source which may
make stellar lens events harder to detect. Relative rates may also be
affected by the orientation of Macho orbits. The Macho velocities in
Table~\ref{tab-rate} assume an isotropic velocity distribution,
whereas Machos on radial orbits are expected to produce a larger event
rate, and an enhanced near-far asymmetry \citep{ker03}.

Perhaps surprisingly, the total number of events predicted for the
three models is similar.  The ratio of the total event rates for the
massive halo, spheroid and disk models is 1:1.1:0.9, respectively,
slightly different from the 1:0.9:0.7 ratio for the total numbers of
ongoing events computed from $\tau_{\rm p}$. The difference is due
mainly to the fact that stellar lenses typically have smaller
$t_{1/2}$ durations, so models with larger stellar lens contributions
show an increase in the rate relative to their optical depth
contribution. Overall the total number of events is a remarkably
robust quantity given the large differences between the models.

\section{Discussion}\label{discuss}

One of the key differences between M31 and our own Galaxy is that M31
possesses a more massive and luminous stellar spheroid. We have assessed
the consequence of this for pixel-lensing experiments which are
targeting M31. These experiments are hoping to confirm the existence
of massive compact halo objects (Machos) by detecting an excess of
events towards the far side of the M31 disk. The near-far spatial
asymmetry expected for a spheroidal distribution of M31 Machos is
regarded as a crucial test of the Macho hypothesis.

In this study we have shown that the presence of a massive stellar
spheroid can induce asymmetry in the spatial distribution of stellar
lenses. Analytical predictions, confirmed by more detailed model
calculations, show that stellar lenses in the disk and spheroid are
asymmetrically distributed if the spheroid and disk mass-to-light
$(M/L)$ ratios differ. A higher spheroid $M/L$ should result in a
stellar lens asymmetry in the same direction as that of Machos
(i.e. an excess of far-disk events), whilst a higher disk $M/L$
results in an asymmetry of opposite sign to the Macho asymmetry
(i.e. an excess of near-disk events). A positive consequence of this
is that two competing dark matter scenarios, one in which one has a
spheroidal dark Macho halo and one in which the dark lenses occupy a
massive disk or highly flattened halo, should be distinguishable by
the fact that they predict spatial asymmetries of opposite sign.

We have constructed three extreme models of M31, each comprising disk,
spheroid and Macho halo populations. The massive halo model assumes
conventional $M/L$ values for the spheroid and disk whilst the massive
spheroid and disk models assume high $M/L$ values consistent with dark
matter dominated stellar populations. By defining the pixel-lensing
optical depth we have computed the number of ongoing events for each
population and have constructed microlensing maps over a region of M31
currently being monitored by a number of survey teams.  Whilst Machos
make up $90\%$ of the events in the massive halo model, stellar lenses
provide at least half of the signal for the other two models, even for
a full Macho halo. However, even for these extreme stellar models, M31
Machos should dominate over most of the far disk if their halo
fraction is around $25\%$ or larger.

We find for all three models that the inclusion of the spheroid
dilutes the Macho near-far asymmetry signature by about $20-30\%$ over
the disk-only asymmetry values. The overall asymmetry varies from 1.4
and 1.7 for the massive disk and spheroid models to 3.4 for the
massive halo model. \citet{ker03} argue that asymmetries weaker than a
factor of 2 will be difficult to detect with current surveys, so if
the M31 disk and spheroid is strongly dark matter dominated (in
contrast to our own Galaxy) they could conceivably mask the asymmetry
signature due to Machos. However, from general considerations of the
likely Macho and stellar lens velocities we argue that many stellar
lenses have durations too short to be detected by the survey teams. If
Machos have a mass of around $0.5\,\sm$, as favoured by the MACHO LMC
survey \citep{alc00} then the overall asymmetry of detected events is
likely to be significantly larger than these estimates because the
longer duration Macho events are easier to detect. Estimates of the
expected event rates indicate that the total number of pixel-lensing
events is remarkably insensitive to the relative mass normalisation of
the halo, spheroid and disk components, though, again, there is likely
to be a strong bias towards the detection of stellar-mass Machos over
stellar lenses because of the short duration of stellar lensing
events.

In summary, the presence of the spheroid complicates matters for
pixel-lensing surveys of M31 and cannot be ignored. A positive
detection of near-far asymmetry cannot, by itself, be taken as a
confirmation of the Macho scenario since a stellar spheroid will
produce a similar result if its $M/L$ is larger than that of the
disk. In this case an analysis of the spatial extent is likely to be
able to distinguish Machos from spheroid lenses, as spheroid lenses
should be more strongly concentrated toward the M31 centre.
The spheroid also has beneficial consequences for the microlensing
survey teams. Candidate samples may be dominated by Machos, or by
lenses in a highly flattened halo or disk, or they may instead be
hoplessly contaminated by variable stars mistaken for
microlensing. In the absence of a spheroid it might prove very difficult to distinguish between the latter two cases. The presence of the spheroid
provides an all-important third signature with which one may be able to distinguish between all three scenarios. In the Macho case one expects a near-far asymmetry. In the case of a highly flattened lens population one expects a far-near asymmetry
(i.e. an asymmetry of opposite sign to that of Machos) due to the presence of the spheroid. Finally, if
the candidates are mostly variable stars their distribution should be
symmetric.


\begin{thebibliography}{}

\bibitem[\protect\citeauthoryear{Afonso et al.}{2003}]{afo03}
Afonso C. et al., 2003, A\&A, 400, 951

\bibitem[\protect\citeauthoryear{Alcock et al.}{2000}]{alc00}
Alcock C. et al., 2000, ApJ, 542, 281

\bibitem[\protect\citeauthoryear{Auri\`ere et al.}{2001}]{aur01}
Auri\`ere M. et al., 2001, ApJ, 553, L137

\bibitem[\protect\citeauthoryear{Bahcall \& Soneira}{1980}]{bah80}
Bahcall J., Soneira R., 1980, ApJS, 44, 73

\bibitem[\protect\citeauthoryear{Baltz, Gyuk \& Crotts}{2003}]{bal03}
Baltz E., Gyuk G., Crotts A., 2003, ApJ, 582, 30

\bibitem[\protect\citeauthoryear{Bellazzini et al.}{2003}]{bel03}
Bellazzini M. et al., 2003, A\&A, 405, 867

\bibitem[\protect\citeauthoryear{Cr\'ez\'e, Chereul, Bienaym\'e \& Pichon}{1998}]{cre98}
Cr\'ez\'e M., Chereul E., Bienaym\'e O., Pichon C., 1998, A\&A, 329, 920

\bibitem[\protect\citeauthoryear{Crotts}{1992}]{cro92}
Crotts A., 1992, ApJ, 399, L43

\bibitem[\protect\citeauthoryear{Crotts et al.}{2001}]{cro01}
Crotts A. et al., 2001,
in Menzies J.W., Sackett P.D., eds., ASP Conference Series Vol. 239,
Microlensing 2000: A New Era of Microlensing Astrophysics.
Astron. Soc. Pac., San Francisco, p.318

\bibitem[\protect\citeauthoryear{Durrell, Harris \& Pritchet}{2001}]{dur01}
Durrell P, Harris W., Pritchet C., 2001, AJ, 121, 2557

\bibitem[\protect\citeauthoryear{Evans \& Wilkinson}{2000}]{evans00}
Evans N.W., Wilkinson M., 2000, MNRAS, 316, 929

\bibitem[\protect\citeauthoryear{Ferguson et al.}{2002}]{fer02}
Ferguson A. et al., 2002, AJ, 124, 1452

\bibitem[\protect\citeauthoryear{Gould}{1996}]{gou96}
Gould A., 1996, ApJ, 470, 201

\bibitem[\protect\citeauthoryear{Gould, Bahcall \& Flynn}{1997}]{gou97}
Gould A., Bahcall J.N., Flynn C., 1997, ApJ, 482, 913

\bibitem[\protect\citeauthoryear{Guhathakurta, Reitzel \& Grebel}{2000}]{guh00}
Guhathakurta P., Reitzel D., Grebel E., 2000, in Discoveries and
Research Prospects from 8- to 10-Meter-Class Telescopes, ed Bergeron
J., Proc. SPIE Vol 4005, p168

\bibitem[\protect\citeauthoryear{Gyuk \& Crotts}{2000}]{gyuk00}
Gyuk G., Crotts A., 2000, ApJ, 535, 621

\bibitem[\protect\citeauthoryear{Holland, Fahlman \& Richer}{1996}]{hol96}
Holland S., Fahlman G., Richer H., 1996, AJ, 112, 1035

\bibitem[\protect\citeauthoryear{Kent}{1989}]{kent89}
Kent S.M., 1989, AJ, 97, 1614

\bibitem[\protect\citeauthoryear{Kerins et al.}{2001}]{ker01}
Kerins E. et al., 2001, MNRAS, 323, 13

\bibitem[\protect\citeauthoryear{Kerins et al.}{2003}]{ker03}
Kerins E. et al., 2003, ApJ, in press

\bibitem[\protect\citeauthoryear{Lasserre et al.}{2000}]{las00}
Lasserre T. et al., 2000, A\&A, 355, L39

\bibitem[\protect\citeauthoryear{Pritchet \& van den Bergh}{1994}]{pri94}
Pritchet C.J., van den Bergh S., 1994, AJ, 107, 1730

\bibitem[\protect\citeauthoryear{Reitzel, Guhathakurta \& Gould}{1998}]{rei98}
Reitzel D., Guhathakurta P., Gould A., 1998, AJ, 116, 707

\bibitem[\protect\citeauthoryear{Riffeser et al.}{2001}]{riff01}
Riffeser A. et al., 2001, A\&A, 379, 362

\bibitem[\protect\citeauthoryear{Walterbos \& Kennicutt}{1987}]{wal87}
Walterbos R., Kennicutt R., 1987, A\&AS, 69, 311

\bibitem[\protect\citeauthoryear{Widrow, Perrett \& Suyu}{2003}]{wid03}
Widrow L., Perrett K., Suyu S., 2003, ApJ, 588, 311

\bibitem[\protect\citeauthoryear{Wielen, Jahreiss \& Kr\"uger}{1987}]{wie87} 
Wielen R., Jahreiss H., Kr\"uger R.,
1987, in The Nearby Stars and the Stellar Luminsoity Function,
Proc IAU Colloq 76, eds A.G. Davis Philip, A.R. Upgren, L. Davis Press: Schenectady, New York, p163


\end{thebibliography}
\end{document}